\documentclass[screen, sigconf]{acmart}

\AtBeginDocument{%
  }
\setcopyright{acmlicensed}
\copyrightyear{2026}
\copyrightyear{2026}
\acmYear{2026}
\setcopyright{cc}
\setcctype{by}
\acmConference[CHI '26]{Proceedings of the 2026 CHI Conference on Human Factors in Computing Systems}{April 13--17, 2026}{Barcelona, Spain}
\acmBooktitle{Proceedings of the 2026 CHI Conference on Human Factors in Computing Systems (CHI '26), April 13--17, 2026, Barcelona, Spain}
\acmPrice{}
\acmDOI{10.1145/3772318.3790365}
\acmISBN{979-8-4007-2278-3/2026/04}

\usepackage{makecell}

\begin{document}

\title{Needling Through the Threads: A Visualization Tool for Navigating Threaded Online Discussions}

\newcommand{\system}{\textit{Needle}}

\author{Yijun Liu}
\orcid{0009-0008-6601-9237}
\affiliation{%
  \institution{University of Illinois Urbana-Champaign}
  \city{Urbana}
  \state{IL}
  \country{USA}
}
\email{yijun6@illinois.edu}

\author{Frederick Choi}
\orcid{0000-0002-8818-2456}
\affiliation{%
  \institution{University of Illinois Urbana-Champaign}
  \city{Urbana}
  \state{IL}
  \country{USA}
}
\email{fc20@illinois.edu}

\author{Eshwar Chandrasekharan}
\orcid{0000-0002-7473-1418}
\affiliation{%
  \institution{University of Illinois Urbana-Champaign}
  \city{Urbana}
  \state{IL}
  \country{USA}
}
\email{eshwar@illinois.edu}

\begin{abstract}
Navigating large-scale online discussions is difficult due to their rapid pace and high volume of content. Platforms like Reddit employ ``threads’’ to visually organize parallel discussions, but deep nesting obscures conversation flow. For moderators, this fragmentation compounds the difficulty of following evolving conversations and maintaining context across threads, which limits timely and effective moderation. In this paper, we present Needle, an interactive system that applies visual analytics to summarize key conversational metrics: activity, toxicity, and voting trends over time. Needle provides both high-level overviews and detailed breakdowns of threads, enabling moderators to identify priority areas without reading through entire nested conversations. Through a user study with ten Reddit moderators, we find that Needle provides a practical solution to maintain contextual understanding when navigating threaded discussions. Based on these findings, we propose design guidelines for future visualization-based tools that shape how people consume, interpret, and make sense of large-scale online discussions.
\end{abstract}



\begin{teaserfigure}
\centering
\includegraphics[width=\textwidth]{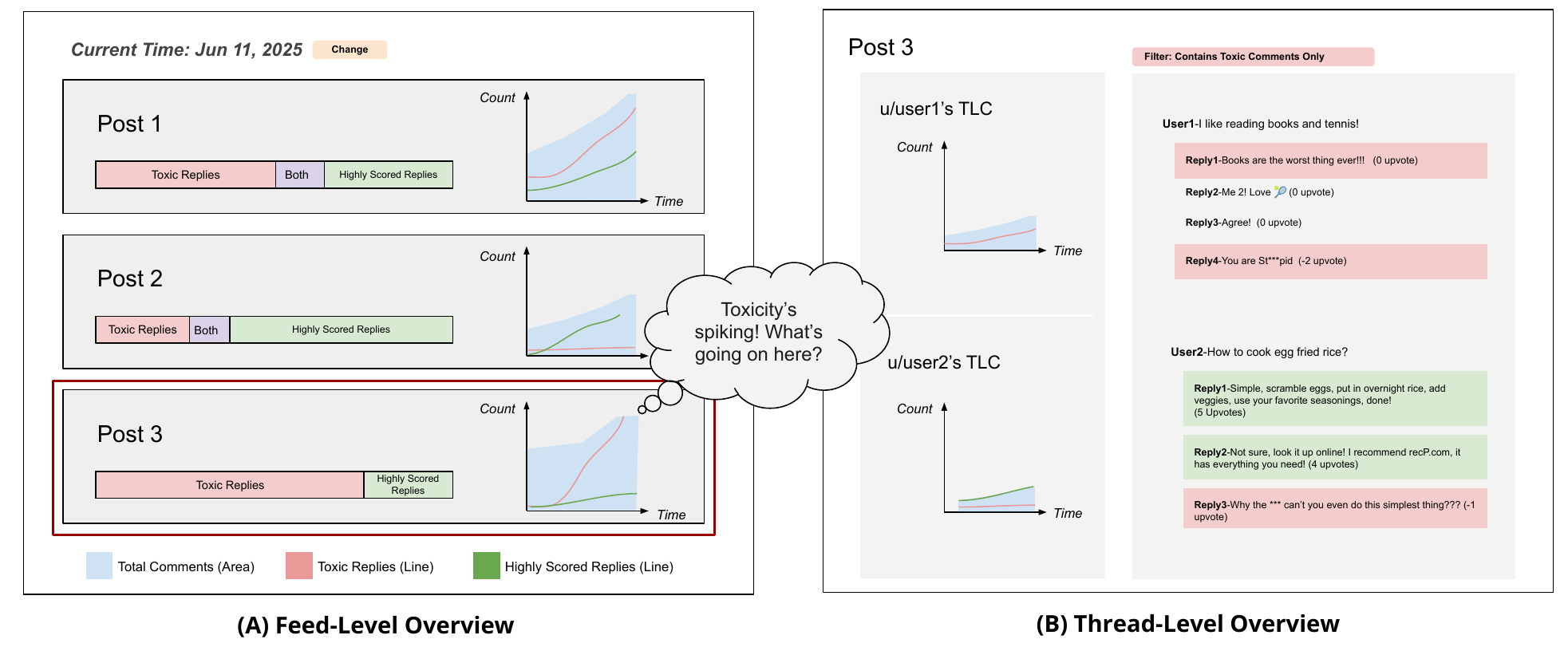}
\caption{\textbf{Needle's Multi-Level Visual Analytics Approach.} (A) The feed-level overview shows temporal visualizations and toxicity patterns across all subreddit posts, with visual anomalies (like Post 3's toxicity spike) drawing immediate attention. (B) The thread-level overview supports drilling down into specific threads with Top-Level Comment (TLC) visualizations as well as highlighting and filtering capabilities to quickly identify problematic content requiring moderator attention.}
\label{fig:needle_overview}
\Description{Needle's visualization has two linked views. The feed-level overview on the left shows bar and temporal charts of comments; bar chats show a break down of comments that are toxic, highly scored, or both; temporal charts are line graphs with time on the x-axis and number of comments in the y-axis; Post 3 is flagged for a sharp rise of toxicity comments. The thread-level overview on the right expands Post 3, displaying the same temporal charts but now they are for each top-level comment, and the actual comments on the right side, with toxic replies highlighted in red and constructive replies in green.}
\end{teaserfigure}

\maketitle


\section{Introduction}
Online communities and their members are frequently exposed to \textit{undesirable} content (e.g., toxicity and harassment) which can reduce user perceptions of quality and safety~\cite{twiterToxic, darkFacebook, defineToxicity}, and sometimes even lead to negative psychological effects~\cite{journalistHarm, saha2019prevalence}. To combat online misbehavior, content moderation is the primary mechanism employed by social media platforms~\cite{grimmelmann2015virtues, moderationVis, onlineMentalHealthModeration}. 
Most online communities employ the services of human moderators (either paid or volunteer, centralized or distributed) to regulate user-generated content~\cite{gillespie2018custodians, roberts2019behind}. Prior work has highlighted the numerous challenges faced by human moderation approaches~\cite{gillespie2018custodians, roberts2019behind, seering2019moderator, emotionalLabor}, including the substantial impact of constant exposure to toxic content on the mental health of moderators~\cite{emotionalLabor, CaseyModerationChallenge}. 

To alleviate some of these challenges, prior work in HCI has developed computational tools and approaches to augment the process of manual review by moderators~\cite{onlineHarrLitReview, taylor2018watch, jhaver2019explanations, matias2018civilservant} and support human-in-the-loop approaches to content moderation~\cite{chandrasekharan_crossmod_2019, convex, nipitinthebud}.
Despite the increasing adoption of human-AI systems for content moderation---particularly in light of recent advances in NLP and the emergence of large language models (LLMs)---automated approaches continue to face limitations, including poor understanding of varied community norms~\cite{chandrasekharan_crossmod_2019}, lack of human precision to fully address complex cases~\cite{llm-mod}, and diminished transparency in moderation processes that obscures how and why decisions are made~\cite{politicsAlgo}.

\subsection{Complexities in Navigating Threaded Online Discussions}
Large-scale online discussions further exacerbate moderation challenges due to their volume and the rapid pace of content generation~\cite{CaseyModerationChallenge, chan2022community, chan2024understanding, chan2025examining}.
The scale and complexity of online discourse make it difficult to track conversations and maintain contextual understanding of the surrounding discussions, hindering timely and effective moderation~\cite{glaser2018,jhaver2019automated,madrigal2018}. 
For instance, a recent study highlighted how Discord moderators struggle to monitor multiple discussions that take place in parallel on different channels, track how individual conversations flow or evolve, all while maintaining contextual understanding of the surrounding discussion~\cite{convex}.

\begin{figure*}[t]
    \centering
    \includegraphics[width=\textwidth]{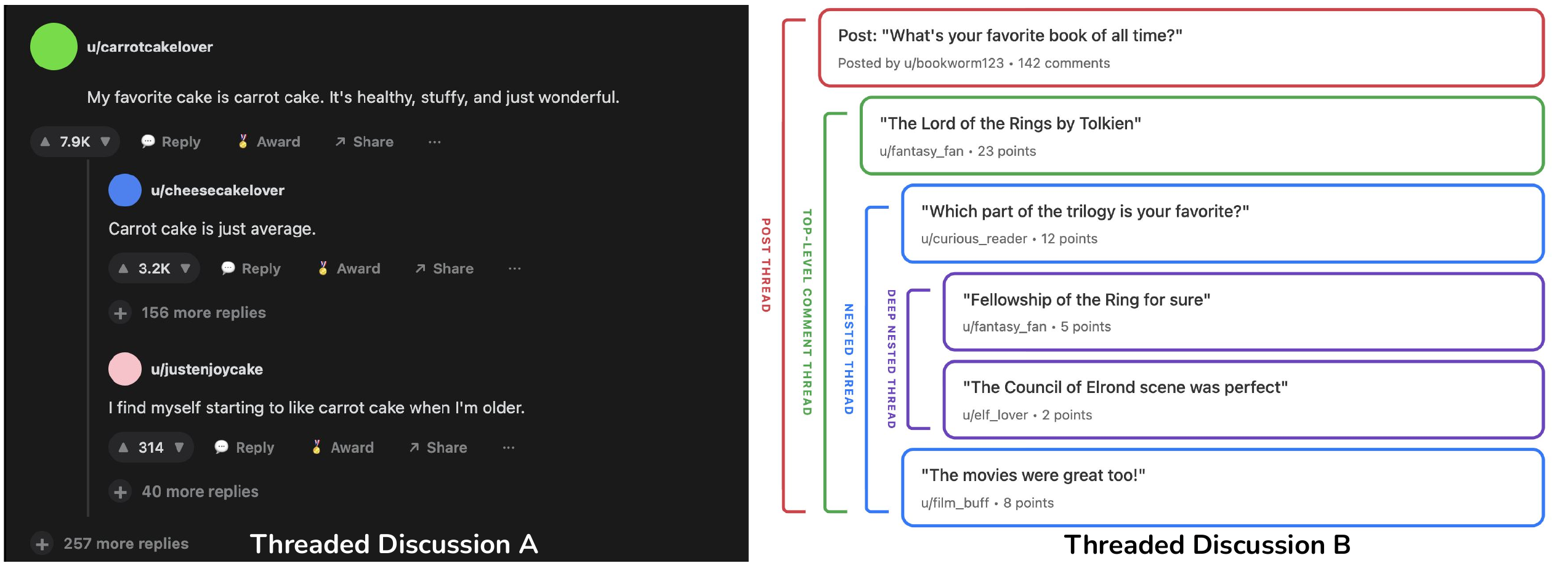}
    \caption{Threaded Discussion A is a mock discussion designed to reflect the scale and comment volume of a real large-scale Reddit thread.
    Threaded Discussion B illustrates the ``threads'' that occur at multiple levels within online discussions using synthetic data.}
    \label{fig:thread-define}
    \Description{The figure shows two examples of threaded discussions. Discussion A, on the left, shows a mock discussion designed to reflect an actual Reddit conversation with more than 250 replies branching from top-level comments. Discussion B, on the right, is a schematic diagram using synthetic data that labels the hierarchy of threads: post thread, top-level comment thread, nested thread, and deep nested thread.}
\end{figure*}

Platforms like Reddit employ ``threads'' to visually organize or structure parallel discussions. Threads can occur at different levels, from posts to comments, as shown in Figure~\ref{fig:thread-define} (see Threaded Discussion A), fragmenting context across multiple layers. Threaded structures offer a degree of organization by allowing discussions to branch logically, which can help users follow discussion paths with greater ease~\cite{AmyFormFrom, budak_threading_2017}. However, as discussions become increasingly nested (i.e., ``threaded''), maintaining context and identifying key points across multiple layers becomes challenging~\cite{gurjar2025argcmv}, as illustrated in Figure~\ref{fig:thread-define} (see Threaded Discussion B).
Threaded structures fragment conversations across multiple layers, making it difficult to maintain local context in large-scale discussions, especially when topics have drifted \cite{Park2016TopicDrift, aragon_thread_2017}. 
Prior work has shown that greater structural complexity reduces overall comprehensibility and readability, and that threaded formats can make it harder for readers to locate the most recent comments, creating additional navigation burdens for moderators who must track evolving conversations across multiple branches \cite{aragon_thread_2017, hadfi_structural_2024}. For moderators, this fragmentation compounds the difficulty of tracking evolving conversations across branches and identifying problematic contents, showing a need for systems that help moderators navigate complex threaded discussions that do not require reading through entire nested conversations \cite{convex}.

HCI researchers have explored tools to enhance sense-making, summarization, and visualization of large-scale online discussions in different contexts~\cite{AmyFormFrom}.
Specifically, text summarization/synthesis and visualization techniques have shown promise in making sense of complex discussions forums and Wikis~\cite{AmyWikum} and conversations on Slack~\cite{AmyTilda}.
In the context of content moderation, previous systems have demonstrated that visual cues can improve user awareness when interacting with \textit{stranger} accounts on Twitter~\cite{synthesizedSocialSignals}, in addition to highlighting how visual summaries can guide moderation attention to problematic conversations on Discord~\cite{convex}.
We extend this line of work on using visualizations to assist content moderation in the context of single-thread conversations (or sequential chat) and develop a visualization tool to navigate a new format of online discourse---\textit{threaded} discussions on Reddit. Our system provides high-level syntheses of key metrics to guide moderators’ attention through visualizations, while also overlaying visual cues directly on the conversations to improve contextual understanding and comprehension of deeply nested conversations.

\subsection{Summary of Contributions}

In this paper, we present \textit{Needle}, an interactive system that provides high-level synthesis via visual analytics and aims to support the navigation of threaded discussions on social platforms like Reddit. \system{} can visualize key comment metrics including activity, toxicity, and number of upvotes or downvotes, as well as temporal trends of how these metrics change over time. \system{} is designed to support moderators by providing intuitive and interactive line graph summaries of recent commenting activity and toxicity levels across different threads within a given post (or Reddit submission). Upon clicking a post, moderators can delve into a detailed visual breakdown of comment activity in each Top Level Comment (TLC) as well as the comments themselves.

Through a user study with Reddit moderators ($N=10$), we find that \system{} can help moderators easily navigate threaded discussions in their communities, monitor key conversational metrics, and quickly identify areas of concern within threads. The use of visualizations in \system{} enables moderators to maintain a clear overview of parallel conversations within threads, making it easier to manage complex threaded discussions and prioritize content that requires immediate moderator attention.
\system{}'s design provides a practical solution that helps moderators maintain contextual understanding of surrounding comments in the conversation when reviewing comments within threaded discussions, without increasing cognitive load and immediate exposure to toxic content.

The contributions of this paper are three-fold. 
First, we make a systems contribution by developing a novel sociotechnical system to help Reddit moderators navigate threaded online discussions. Unlike prior work that primarily targeted linear conversations such as Slack or Discord, our system addresses the unique fragmentation and complexity of nested threads. \system{} is one of the first systems to combine interactive visual analytics with \textit{human-in-the-loop} moderation for threaded online conversations.

Second, we examine how well design guidelines synthesized from prior HCI work on content moderation, visualization, and sociotechnical systems translate into practice.
We develop \system{} based on these design guidelines and evaluate both the guidelines and resulting design features through a user study with ten moderators. 
In addition to validating the usability and practical viability of \system{}, we present insights into which features proved particularly useful and how \system{} can be improved.
Finally, we present additional guidelines and considerations for designers of future tools to navigate and summarize online discussions using visualizations, especially when threaded conversations are involved.
Our broader goal is to inform the design of visualization-based tools that can guide how individuals process, understand, and make sense of large-scale online discussions. 

\section{Related Work}

In this section, we review prior literature to summarize key challenges in human moderation approaches and situate our work in the context of recent advances in computational approaches to content moderation and related visual interfaces for navigating large-scale online discussions.

\subsection{Challenges in Human Moderation}
Online harassment, toxicity, and other forms of harmful content present serious risks to user well-being and community health \cite{onlineHarrLitReview, community_rules}, making content moderation a critical function for digital platforms~\cite{grimmelmann2015virtues}. As highlighted by prior research, human moderation remains the primary approach for managing harmful content across these platforms~\cite{moderationVis, onlineMentalHealthModeration, gillespie2018custodians}. But the increasing amounts of manual labor performed by human moderators has the potential to result in severe cognitive and emotional strain~\cite{roberts2019behind}. Frequent exposure to toxic content and manually reviewing large volumes of user-generated content has been linked to detrimental impact on moderators, including negative outcomes like moderator burnout, stress, and ``secondary trauma''~\cite{CaseyModerationChallenge, emotionalLabor, psycModerator}. These challenges are especially pronounced in volunteer-driven communities~\cite{kiene2019volunteer} like Reddit~\cite{li2022all}, Discord~\cite{CaseyModerationChallenge}, and Twitch~\cite{cai2021after}, where only a few moderators need to perform thousands of actions every day~\cite{habib2019actreactinvestigatingproactive, emotionalLabor}.

The scale of user-generated content presents an additional challenge~\cite{glaser2018,jhaver2019automated,madrigal2018}, with major platforms such as TikTok receiving millions of new posts daily~\cite{inserra2024guide}. This volume renders purely manual review impossible and has led to an increasing reliance on computational approaches to assist human moderation.

\subsection{Computational Approaches and Tools to Assist Content Moderators}
To scale moderation efforts, platforms and researchers have increasingly developed automated tools and algorithmic approaches to detect norm violations and assist moderator decision-making \cite{politicsAlgo, proactive_review, onlineHarrLitReview, matias2018civilservant}. These computational approaches generally fall into three categories: text-based classifiers and filters \cite{jhaver2019automated, 10.1145/2998181.2998277}, pattern and behavior analysis \cite{saveski_social_2021, yu_signals_2025, yu_characterizing_2024, noauthor_spreading_nodate, cheng_anyone_2017, saveski_structure_2021, diakopoulos_towards_2011}, and the use of large and small language models \cite{llm-mod, zhan2024slm, goyal2025momoe}. Text-based classifiers and filters are used to identify formatting violations and filter content based on lists of predefined keywords \cite{jhaver2019automated, 10.1145/2998181.2998277}. For example, Reddit employs a built-in rule-based tool called AutoModerator (AutoMod).\footnote{https://www.reddit.com/r/reddit.com/wiki/automoderator/}
Structural approaches analyze engagement patterns. Prior work indicates that toxic conversations and patterns of how misinformation cascades exhibit distinct structural patterns \cite{saveski_structure_2021, noauthor_spreading_nodate}. Prior work demonstrates that identifying deviations from predicted engagement levels \cite{yu_signals_2025} and analyzing the topology of discussion threads \cite{yu_characterizing_2024} can help characterize and understand anomalies in conversations, in addition to pure text-based metrics. Furthermore, these structures capture situational dynamics, such as the influence of ``social catalysts'' (e.g., users who ignite conversations) \cite{saveski_social_2021} or trolling behaviors due to surrounding discussions \cite{cheng_anyone_2017}.

Yet these systems remain limited by specific trade-offs. Structural signals often require a conversation to mature before patterns emerge \cite{saveski_structure_2021}, limiting their utility for immediate, real-time detection. Text-based filters often fail in their ability to interpret nuance, handle linguistic subtleties like sarcasm or coded language, and understand community-specific norms \cite{thread_caution, politicsAlgo, proactive_review}. For example, fully automated filters may mislabel heated but productive debate as toxic, or overlook veiled harassment masked in polite phrasing \cite{conversations_gone_awry}.
As a result, automated moderation systems continue to struggle with context-dependent nuance~\cite{chandrasekharan2018internet} and cultural variations \cite{politicsAlgo}---even with recent approaches that rely on large language models \cite{llm-mod}.

These limitations have prompted calls for human-in-the-loop systems that blend computational efficiency with human discernment \cite{chandrasekharan_crossmod_2019, convex, nipitinthebud, halfaker2020ores}. Such systems do not aim to replace human moderators but to enhance their capabilities by surfacing high-risk content for review, visualizing user activity patterns, or highlighting contextual signals \cite{convex, proactive_review}. For example, ConvEx~\cite{convex} utilizes visualizations and AI to guide moderator attention to toxic messages on Discord, and Venire~\cite{koshy2024veniremachinelearningguidedpanel} helps Reddit moderation teams identify reports where moderators are likely to have disagreements. Importantly, research shows that moderators value tools that increase transparency and interpretability, and facilitate informed and community-aligned decision-making \cite{thread_caution}.

\textit{Extending this line of work, we aim to develop a system that supports human-in-the-loop moderation and incorporates computational techniques to assist, rather than replace, human judgment.
}
\subsection{Visual Interfaces for Navigating Large-Scale Conversations}

As platforms like Reddit employ threaded discussions to structure and organize content, these threaded structures introduce complexity as conversations branch into multiple pathways and foster increased social reciprocity between users \cite{AmyFormFrom, aragon_thread_2017}. This branching architecture, although beneficial for organization, generates long, intricate conversation networks that impose substantial cognitive demands on moderators attempting to maintain contextual understanding across discussion threads \cite{kang_thread, aragon_thread_2017}. Consequently, moderators face increasing difficulty in synthesizing the necessary context for informed decision-making as conversations expand in both breadth and depth.

Researchers in HCI have explored visualization methods to address these challenges in making sense and navigating through large online discussions. Early visualization efforts like Thread Arcs provided compact graphical representations of both chronological sequences and hierarchical relationships, improving users' ability to grasp thread structures quickly \cite{Kerr_email_thread}. More recent visualization approaches have emphasized multi-scale and interactive timelines. For instance, DisVis employed timeline and post activity graphs combined with detailed summary statistics, allowing users to rapidly identify periods of intense activity or topic shifts \cite{nakikj2017disvis}. 

Similarly, analytic systems like ConVis \cite{convis} and ConvEx \cite{convex} incorporated linguistic analysis into visual interfaces, helping users quickly pinpoint emerging trends and contentious issues within expansive discussions. AudienceView leverages NLP and LLMs to analyze and visualize themes, sentiment, and topic clusters from hundreds of thousands of YouTube comments to support journalistic sense-making \cite{10.1145/3678884.3681821}. Additionally, \citet{10.1145/3584931.3607499} uses both concept maps and comment tree maps to help novice learners navigate large video collections through structured content and social discussion themes. Overall, these visual interfaces have consistently demonstrated their ability to reduce cognitive load and the potential to enhance moderator's experience by clearly presenting complex conversational structures and temporal dynamics.

\textit{In this paper, we examine how visualizations can be used to easily navigate and summarize online discussions involving threaded conversations, specifically focusing on Reddit moderators. }

\begin{figure*}[t]
  \centering
  \includegraphics[width=0.85\linewidth]{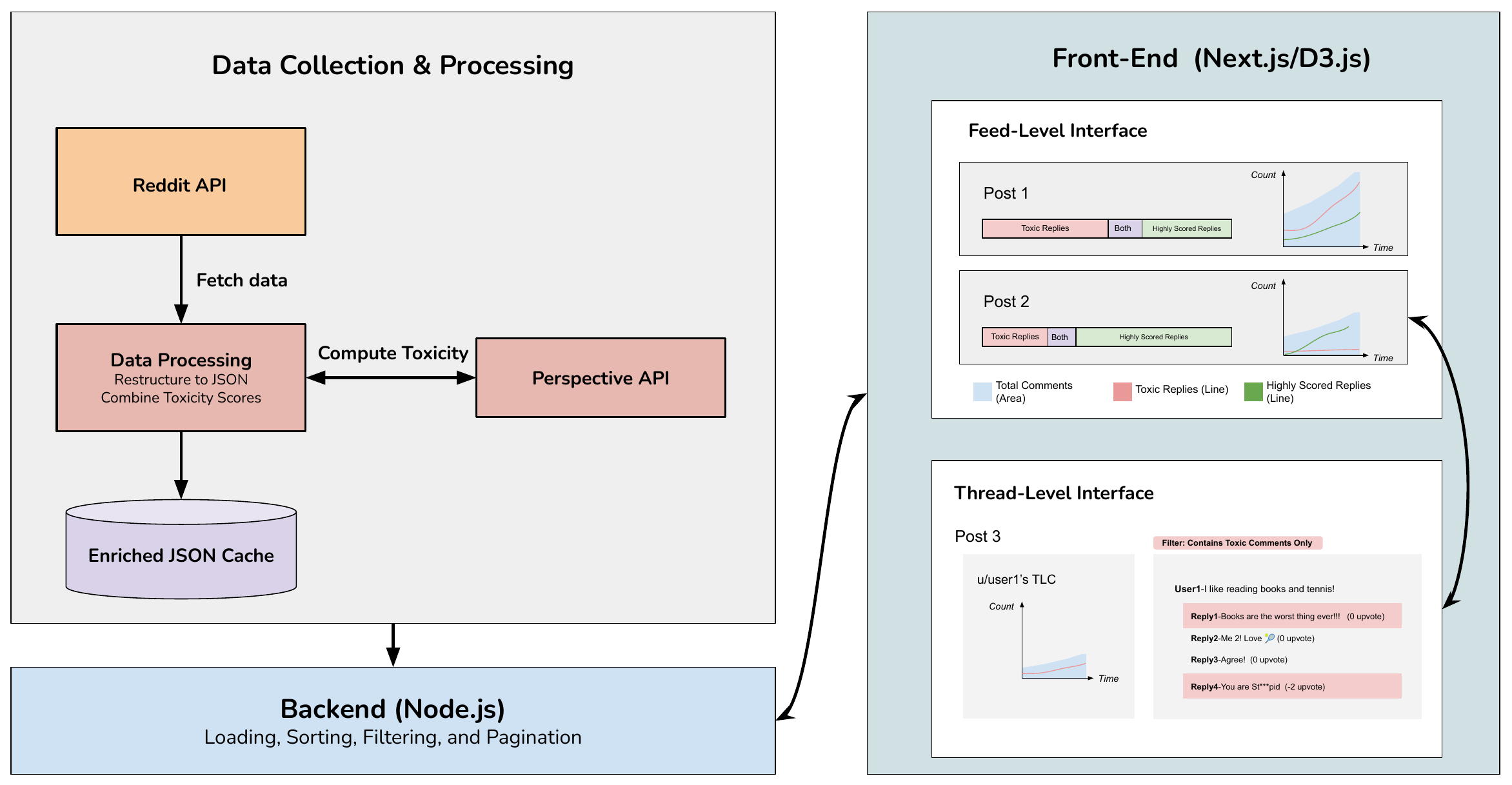}
  \caption{\system{}'s System Architecture. Data from Reddit is processed and augmented with toxicity scores, stored as a JSON cache, and served by a Node.js backend to a Next.js/D3.js frontend; front-end is powered by two pages: Feed-level and Thread-level interfaces.}
  \Description{The system architecture diagram shows three parts: data collection and processing, front-end, and backend. Reddit comments are fetched via the Reddit API, processed into JSON and enriched with toxicity scores using the Perspective API, then stored in a cache. A Node.js backend handles loading, sorting, and filtering before sending data to a Next.js/D3.js frontend. The frontend has two views: a feed-level interface and a thread-level interface.}
  \label{fig:pipeline}
\end{figure*}

\section{Design Goals}
\label{sec:design-goals}
Next, we present the three primary design goals that guided the development of our new visual analytics system to enhance moderation on Reddit, a threaded discussion platform. These design goals were derived from a comprehensive review of prior literature on content moderation and the specific needs of moderators working with large-scale online discussions on platforms like Reddit.

\subsection{Supporting Multi-level Summarizations through Visual Aids}

Research on content moderation consistently shows that moderators struggle to maintain awareness across multiple discussions simultaneously \cite{CaseyModerationChallenge, emotionalLabor}. The sheer volume and complexity of online discussions often overwhelm a moderator's capacity to efficiently identify areas requiring intervention, leading to delayed interventions on problematic content and moderator burnout \cite{emotionalLabor}. Previous moderation tools have typically focused on content flagging and reporting mechanisms, which provide limited contextual information. Visualization approaches like those explored by \citet{convex} have shown promise in reducing cognitive load through visual summarization, but these lack the multi-level perspective needed for complex threaded discussions. 

Information visualization theory suggests that effective overviews combined with details-on-demand can significantly improve situation awareness for complex monitoring tasks \cite{1509067}. We apply this in the context of online moderation, supporting moderator awareness by surfacing trends in recent activity through multi-level visual summaries.

\textbf{Design Goal 1 (G1): \system{} should provide multi-level visual summaries (subreddit and thread levels) and surface trends in recent activity to support efficient monitoring and improve moderator awareness.}

\subsection{Directing Attention and Providing Actionable Insights}

With limited time and cognitive resources, moderators benefit from tools that help prioritize their attention and provide contextual information for decision-making \cite{moderationVis}. However, existing approaches fail to provide sufficient support for moderators. While community reports and flagging systems provide the primary mechanism for helping moderators prioritize their attention, they rely on community members to identify and report problematic content \cite{crawford_what_2016} and flagged content appears in a separate interface (e.g., Reddit's ModQueue). This approach can miss subtler issues or novel forms of policy violations that community members do not recognize or report, as well as showing flagged content in an isolated queue may limit moderators' understanding of the conversations trends \cite{bajpai_queue_2025}. Previous approaches have also explored automated content detection using machine learning \cite{politicsAlgo}, but these often lack contextual awareness and have high false positive rates when used alone. 

However, research on attention guidance in visual analytics suggests that combining algorithmic signals with community indicators and temporal patterns can more effectively guide moderator attention \cite{convex}. We follow this approach to support informed decision-making by directing moderator attention to key discussion and comments.

\textbf{Design Goal 2 (G2): \system{} should direct moderators' attention to key discussions and comments using metrics and visual cues that support informed decision-making.}

\subsection{Adapting to Diverse Community Needs}

Online communities vary widely in their norms, sizes, and content policies \cite{psycModerator}. One-size-fits-all approaches to moderation fail because what constitutes problematic content in one community may be acceptable or even encouraged in another \cite{jhaver2019automated}. However, many current moderation systems typically offered limited customization options, forcing communities to adapt to tools rather than adapting tools to communities \cite{jhaver2019automated}.

Moderators need tools that can adapt to these differences without requiring technical expertise to configure them. Additionally, discussion sizes range from small exchanges to massive threads with thousands of comments, requiring scalable approaches to visualization and analysis. Research on end-user customization suggests that providing moderators with accessible controls can improve adoption and usages \cite{grimmelmann2015virtues, kiene_technological_2019, convex}. Visualization research also indicates that adaptive displays that respond to data characteristics can better maintain usability across varying scales of information \cite{shneiderman_eyes_nodate}. We thus aim to provide customization options that allow moderators to adapt the tool to the needs and characteristics of their specific communities.

\textbf{Design Goal 3 (G3): \system{} should adapt to different community needs and discussion sizes through customizable settings.}


\section{\system{}: A Visualization Tool to Navigate Threaded Discussions} 

In this paper, we present \system{}, a sociotechnical system designed to help Reddit moderators navigate, interpret, and act on complex threaded discussions.\footnote{Code is available in the following repository: \url{https://github.com/scuba-illinois/Needle}} 
\system{}'s development includes a data processing pipeline and two interfaces: a feed-level overview (Figure~\ref{fig:interface}) that summarizes activity and toxicity across posts, and a thread-level drill-down (Figure~\ref{fig:interface-thread}) that supports detailed inspection of individual threads. In this section, we describe the system architecture, interfaces, and visualizations, as well as show how specific features within them support each specific design goal (Section~\ref{sec:design-goals}): multi-level summarization (G1), attention guidance (G2), and community customization (G3).

\subsection{System Architecture}
\system{} is implemented as a client-side web application using Next.js for the frontend, D3.js for interactive visualization, and Node.js for the server-side API layer that serves preprocessed data. In this section, we will walk through the details of implementing \system{}. The complete system architecture is shown in Figure~\ref{fig:pipeline}.

\subsubsection{Data Collection \& Processing} 
\label{sec:data_collection}
We retrieved posts and comments from the Reddit API. For each post and comment, we extracted metadata including \texttt{id}, \texttt{author}, \texttt{text}, \texttt{timestamp}, and \texttt{score}. Score, obtained directly from the Reddit API, is roughly the net total of upvotes minus downvotes displayed next to the voting arrows. It is the main mechanism for users to affect comments' visibility in Reddit \cite{horne_identifying_2017}. Comments were recursively expanded to preserve the threaded structure. This structure allows the frontend to preserve post $\rightarrow$ top-level comment $\rightarrow$ reply hierarchies in the threaded nature.

In addition to data from the Reddit API, we also leveraged the Perspective API \cite{lees_new_2022} to assess toxicity levels in comments—specifically, we used the TOXICITY model.\footnote{The Perspective TOXICITY model is a machine-learning classifier trained on online comments. Its architecture consists of multilingual BERT-based models distilled into language-specific convolutional neural networks \cite{perspective_model_card}.} This tool provides a continuous score indicating the likelihood that a comment may be perceived as toxic, ranged from 0 to 1, as 1 being the most toxic. The Perspective API is widely used in HCI and NLP research \cite{convex, yang_towards_2023, raman_centering_2023, zheng_hatemoderate_2024} and serves as a validated baseline for toxicity detection. It provides a standardized and interpretable score, supports direct comparability with prior moderation studies, and avoids the computational and ethical challenges of training custom models that solely rely on removed content.

\subsubsection{Backend}

The backend is implemented in Node.js and acts as a lightweight server layer between the static JSON data cache and the frontend. Its primary responsibilities are to load the cached dataset, apply client-side requests for sorting, filtering, and pagination, and return results efficiently. 

Unlike systems that recompute toxicity or restructure comments on demand, our backend is deliberately minimal: it only serves preprocessed data to ensure low latency, reproducibility, and robustness under study conditions. This design allows us to control for variability in the Reddit API (e.g., rate limits, changing comment states) while still supporting interactive exploration on the frontend.

\subsubsection{Frontend}
The frontend is implemented in Next.js with D3.js for interactive visualizations. Its role is to render preprocessed data served by the backend. The next section provides a walkthrough of the frontend interfaces and features.

\begin{figure*}[t]
    \centering
    \includegraphics[width=\textwidth]{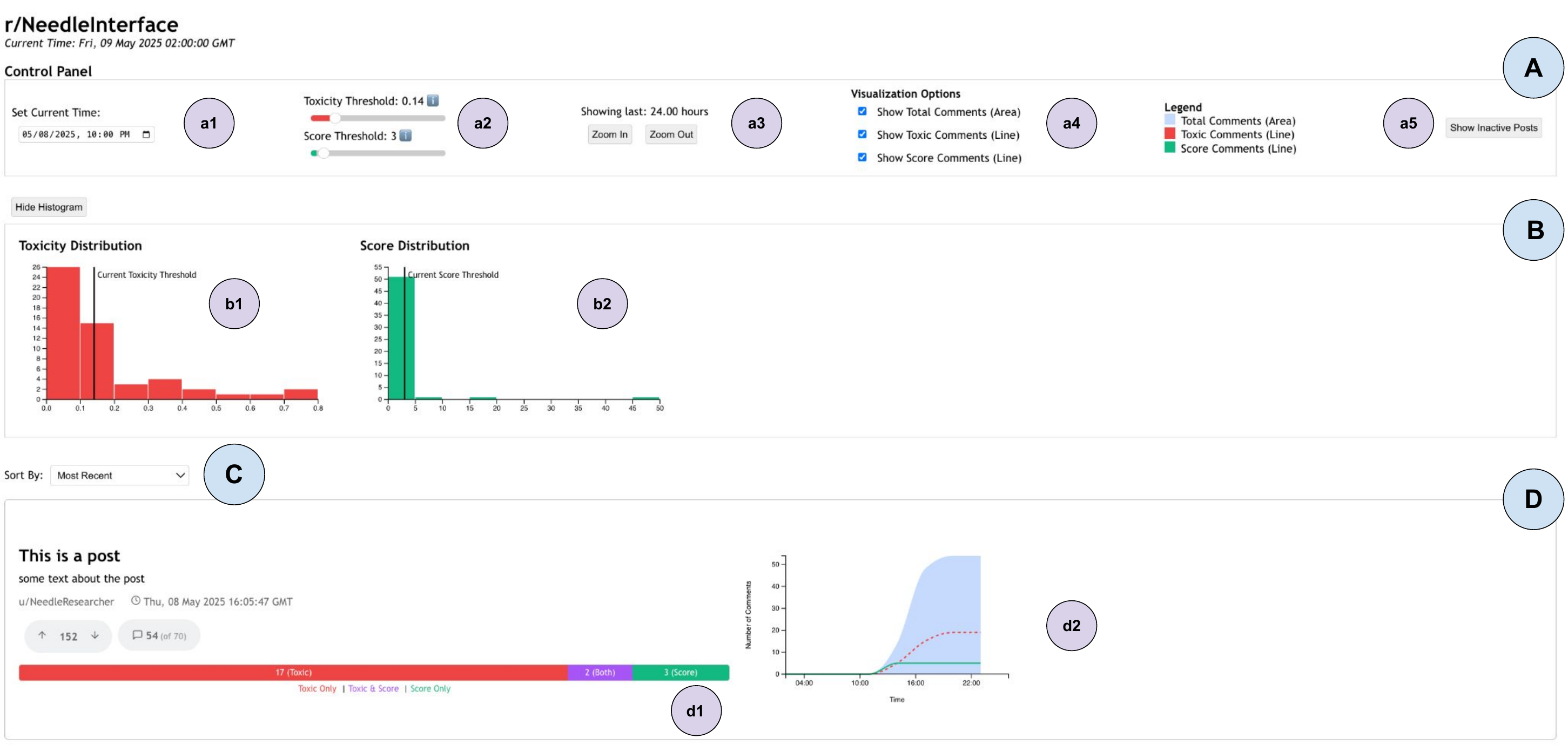}
    \caption{The feed-level interface of \system{}, which presents subreddit-wide comment activity and toxicity. (A) Control panel for setting time (a1), toxicity and score threshold (a2), zoom level for time range (a3), line toggles with legend (a4), and toggle show inactive posts (a5). (B) Global histograms for toxicity (b1) and score distributions (b2). (C) Sorting dropdown to organize posts by criteria. (D) Post summaries, with a segmented comment breakdown bar (d1), and a temporal activity chart (d2).}
    \Description{Screenshot of the feed-level interface of Needle. At the top is the control panel labeled A. Below are global histograms showing distributions of toxicity and scores labeled B. A sorting dropdown appears next labeled C. The lower half shows post summaries (labeled D), each with a segmented bar chart for comment breakdown and a temporal activity chart. The layout flows from top to bottom, with elements arranged left to right within each section.}
    \label{fig:interface}
    
\end{figure*}

\begin{figure*}[t]
    \centering
    \includegraphics[width=\textwidth]{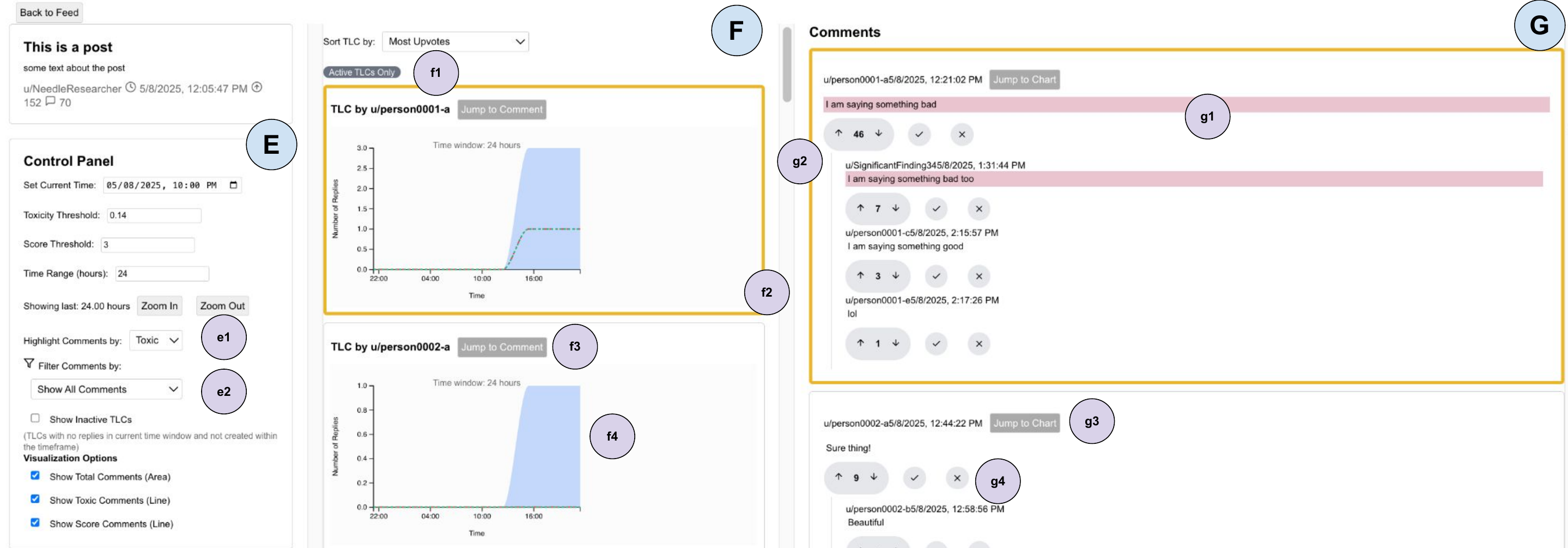}
    \caption{The thread-level interface of \system{}, which provides detailed visualization of individual discussion threads. (E) Control panel with the same customization options as the feed-level interface, adding two new features of applying highlights to comments (e1) and filter comments (e2). (F) Comment visualization features including: active TLCs only indicator (f1), hover effects to indicate linkage between visualizations to comments (f2), jump to Comment button for quick navigation (f3), and temporal activity chart for the selected TLC (f4). (G) Actual comments: toxic comment highlighting (g1), hover effects (g2), navigation controls to move between charts and comments (g3), and moderation buttons (g4).} 
    \label{fig:interface-thread}
    \Description{Screenshot of the thread-level interface of Needle. The layout has three main panels arranged left to right. On the left is the control panel with customization options labeled E. In the center are visualizations of thread activity labeled F. On the right is the comment section itself, showing highlighted toxic comments, navigation controls, and moderation buttons labeled G.}
\end{figure*}

\subsection{Interfaces and Features}
\system{} provides two levels of interfaces aligned with Reddit’s existing structure: a feed-level overview (Figure~\ref{fig:interface}) and a thread-level drill-down (Figure~\ref{fig:interface-thread}). Below we introduce these two interfaces and our design choices, as well as alignment to our design goals.

\paragraph{Metric Selection Rationale} In \system{}, we use toxicity, score, and activity as the key metrics to visualize and guide moderators’ attention. Toxicity captures hostile or harassing language (e.g., insults and threats), which is strongly correlated with removals and community sanctions \cite{vargo_deciding_2024, blumer_tracking_2025, cima_taming_2024}. We prioritized toxicity specifically because it addresses the challenge of scale: in high-volume discussions where exhaustive reading is impossible, toxicity scores help surface problematic content that community reports often miss. Score (upvotes minus downvotes, voting trends) reflects community sentiment on whether a comment is norm-aligned or not, and provides an indirect signal of removal likelihood \cite{graham_sociomateriality_2021, lampe_slashdot_2004, guo_throwaway_2025, vargo_deciding_2024}. Activity measures overall thread engagement (e.g., the total number of comments comes in), providing essential context for interpreting the relative frequency of toxic or highly scored posts within each discussion. 

These metrics serve as interpretative signals rather than absolute determinations. Recognizing that acceptable thresholds for toxicity can vary significantly across communities, \system{} relies on the moderator’s judgment for final adjudication. To facilitate this calibration, the interface includes histograms that visualize the distribution of these scores (Fig.~\ref{fig:interface}B), allowing moderators to quickly establish a baseline and discern what constitutes an ``unusual'' or ``noteworthy'' deviation in their specific community context. To contextualize these metrics, we include Table~\ref{tab:comment_examples} in Appendix~\ref{sec:appendix_examples} that maps specific score and toxicity values to paraphrased example comments from our user studies.

\paragraph{Feed-level interface.}  
The feed-level page (Figure~\ref{fig:interface}) provides an overview of the subreddit and each post, similar to the main subreddit page in Reddit. It summarizes subreddit-wide activity and toxicity at both global and post levels. Global histograms visualize overall distributions of toxicity and scores, while post-level bar charts and temporal activity charts show patterns of toxicity and score within individual posts. 

Moderators may begin with the control panel (Figure~\ref{fig:interface}-A), adjusting thresholds and zoom levels to see how distributions shift across the subreddit; thresholds instantly propagate to post-level views. Hover tooltips clarify metrics like ``toxicity'' and ``score.'' Posts are then browsed (Figure~\ref{fig:interface}-D), by default sorted by recency, but sortable by other criteria, including highest score, highest toxicity, and highest activity. These posts' visualizations can guide attention, for example, a post with a sudden toxic spike becomes more salient for further inspection. Once familiar with \system{}, moderators can leverage additional controls, such as customizing the time range (Figure~\ref{fig:interface}-a3) or toggling specific metrics on and off to refine their view (Figure~\ref{fig:interface}-a4).

\paragraph{Thread-level interface.}  
The thread-level interface (Figure~\ref{fig:interface-thread}) enables detailed examination of individual discussions' comments. This interface uses the temporal visualization approach from the feed-level view, displaying activity patterns within each top-level comment (TLC) thread. Beyond the standard customization controls in the previous page, this page incorporates new functionality including comment highlighting and filtering capabilities, and integrated moderation actions including content approval or deletion to mirror the workflow of Reddit's existing ModQueue interface.

Landing on the page, moderators first see the full set of comments (Figure~\ref{fig:interface-thread}-G) as they occupied the largest area on screen. Toxicity highlights are applied by default (Figure~\ref{fig:interface-thread}-g1). Moderators can then refine the view with filters (Figure~\ref{fig:interface-thread}-e2) such as showing only toxic or high-scoring comments, or change highlight rules (Figure~\ref{fig:interface-thread}-e1). Comment visualizations (Figure~\ref{fig:interface-thread}-F) provide temporal overviews of each TLC (Figure~\ref{fig:interface-thread}-f4), while hover effects (Figure~\ref{fig:interface-thread}-f2) and ``Jump to Comment'' (Figure~\ref{fig:interface-thread}-f3) link the visualizations directly to their source comments. Finally, moderation buttons (Figure~\ref{fig:interface-thread}-g4) allow immediate actions such as removing content.

\begin{table*}[ht]
\centering
\footnotesize
\begin{tabular}{@{}p{3.2cm}p{1.4cm}p{4.4cm}p{3.8cm}@{}}
\toprule
\textbf{Feature} & \textbf{Location} & \textbf{Description} & \textbf{Use Cases} \\
\midrule
\multicolumn{4}{l}{\textbf{Multi-level Summarizations (G1)}} \\
\addlinespace
Toxicity \& score histograms & Figure~\ref{fig:interface}-B & Visualizes the distribution of comment toxicity and score levels across all posts & Assess overall subreddit health and establish appropriate filtering thresholds \\
\addlinespace
Post temporal visualizations & Figure~\ref{fig:interface}-d2 & Charts how discussion activity evolves within individual posts over time & Analyze post engagement patterns and identify toxicity, score, or activity peaks \\
\addlinespace
TLC temporal visualizations & Figure~\ref{fig:interface-thread}-f4 & Charts how discussion activity evolves within individual top-level comment threads over time & Monitor how specific conversation threads develop and escalate \\
\addlinespace[0.3em]
\multicolumn{4}{l}{\textbf{Attention Direction and Actionable Insights (G2)}} \\
\addlinespace
Sort options & Figure~\ref{fig:interface}-C & Reorders posts and top-level comments by various criteria & Prioritize content based on toxicity levels, activity, recency, or scores \\
\addlinespace
Segmented breakdown bars & Figure~\ref{fig:interface}-d1 & Color-coded bars that visually segment comments by toxicity and score numbers & Identify problematic or highly-rated content within posts \\
\addlinespace
Inactive content toggle & Figure~\ref{fig:interface}-a5 & Shows or hides posts and TLCs with minimal recent activity & Reduce interface clutter by focusing on active discussions \\
\addlinespace
Comment highlighting & Figure~\ref{fig:interface-thread}-e1 & Visually emphasizes comments that exceed toxicity or score thresholds & Direct moderator attention to comments requiring immediate review \\
\addlinespace
Comment filtering & Figure~\ref{fig:interface-thread}-e2 & Displays only comments meeting specific toxicity or score thresholds & Concentrate moderation efforts on the most critical content \\
\addlinespace[0.3em]
\multicolumn{4}{l}{\textbf{Community Customization (G3)}} \\
\addlinespace
Threshold configuration & Figure~\ref{fig:interface}-a2 & Allows adjustment of toxicity and score sensitivity levels & Adapt parameters to match different community standards \\
\addlinespace
Time point navigation & Figure~\ref{fig:interface}-a1 & Enables quick jumps to specific timestamps in the activity feed & Review historical activity during specific time periods \\
\addlinespace
Temporal zoom control & Figure~\ref{fig:interface}-a3 & Adjusts the time window scope from 8 minutes to 24 hours & Explore community activities across different temporal scales \\
\bottomrule
\end{tabular}
\caption{Feature overview of \system{} including interface locations, descriptions, and use cases.}
\label{tab:feature-overview}
\Description{Table summarizing the main features of Needle. Features are grouped into three categories: multi-level summarization, focus and attention tools, and community customization. Each entry lists the feature name, its location in the interface, a brief description, and typical use cases.}
\end{table*}

\paragraph{Summary of Features.}
Table~\ref{tab:feature-overview} summarizes the key features of \system{}, grouped according to our three design goals: multi-level summarizations (G1), attention direction and actionable insights (G2), and community customization (G3). Features are listed with their location, function, and representative use cases. This overview illustrates how visual summaries, attention cues, and configurable settings work together to support effective and efficient moderation.

\subsection{Visualizations and Visual Cues}  

While the table above inventories \system{}’s features, here we explain the rationale for our visualization choices for the four types of visualizations we used: temporal visualizations, comment highlighting, segmented breakdown bars, and histograms. Our goal was not simply to display raw data, but to transform the three core metrics---score, toxicity, and activity---into representations that moderators could quickly interpret and act upon.  

Temporal visualizations (Figure~\ref{fig:interface}-d2 and Figure~\ref{fig:interface-thread}-f4) are the key visualizations in \system{} representing multi-level summarization (G1). They were included because activity and pacing are key signals of escalation: bursts of comments or sudden spikes in toxicity often mark moments when intervention is most needed. Raw timestamps are difficult to parse in threaded discussions, but lines and areas in charts allow moderators to perceive these dynamics at a glance.  

Our approach to temporal visualization requires a clear definition of recency within \system{}. Rather than adopting a single fixed threshold, we define recency as a flexible, moderator-controlled window of activity anchored to a chosen point in time. The ``current time'' setting (Figure~\ref{fig:interface}-a1) allows moderators to anchor their analysis at any moment. This anchor pairs with adjustable time range (Figure~\ref{fig:interface}-a3) that scale from minutes to hours, allowing communities with different pacing to calibrate what counts as ``recent'' and to focus on that time frame. 

To balance overview with detail, \system{} visualizes discussions at the feed and Top-Level Comment (TLC) levels. We intentionally do not create separate visualizations for nested replies beneath each TLC, as doing so would introduce excessive visual complexity and cognitive overhead. While the visualization aggregates branches, the comment panel preserves the full hierarchy (Figure~\ref{fig:interface-thread}G). To bridge these views, comment highlighting (Figure~\ref{fig:interface}-g1, that highlights toxicity, score, or both) pinpoints problematic content at any depth, enabling moderators to navigate complex threads without parsing multiple visual layers. Highlighting also draws the eye directly to outliers. We see these design choices as essential to reducing cognitive load and supporting efficient moderation at scale.

Segmented breakdown bars (Figure~\ref{fig:interface}-d1) is another important visualization added to create ``at-a-glance'' profiles of posts. These bars compress all comments into a simple signature, making it easy to notice when a post contains unusually high proportions of toxic or highly scored contributions. This supports triaging: moderators can scan many posts quickly without reading individual comments.  

\begin{table*}[!t]
\centering
\footnotesize
\begin{tabular}{@{}c l l l@{}}
\toprule
\textbf{Participant} & \textbf{Subreddit Size (approx.)} & \textbf{Years Moderating} & \textbf{Subreddit Type} \\
\midrule
P1 & 2300k & 5 & Academic \\
P2 & 400k & 7 & Sports\\
P3 & 700k & 2 & TV shows\\
P4 & 400k & 9 & Games\\
P5 & 600k & 13  & Games \\
P6 & 800k & 8  & Games \\
P7 & 2300k & 9  & Anime\\
P8 & 5900k & 8  & DIY\\
P9 & 4400k & 3 & Sports \\
P10 & 600k & 1  & Anime \\
\bottomrule
\end{tabular}
\caption{Overview of study participants. Subreddit sizes rounded to nearest 100k; years moderating rounded to nearest year.}
\label{tab:participants}
\Description{Table listing ten study participants with their moderated subreddit size, years of experience, and type of subreddit. Subreddit types include games, academic, sports, TV shows, anime, and DIY.}
\end{table*}

In addition to the two key visualizations, we also used histograms (Figure~\ref{fig:interface}-b1 \& b2) to provide necessary context for interpreting toxicity and score as a whole for a subreddit. Instead of providing post-level or thread-level supports, histograms provide a global view of how the subreddit is reacting to the toxicity and score metrics. A toxicity score of 0.7 may be problematic in one community but normal in another; subreddit-wide distributions help moderators calibrate thresholds to their own norms. Without this comparative view, thresholds risk being arbitrary or biased by isolated cases.  

\section{User Study}
Next, we conduct a user study to examine how well the design guidelines synthesized from prior literature translate into practice.
Specifically, we ran an online user study with moderators from various subreddits by simulating a realistic workflow using \system{}.
This allows to examine how \system{} could support Reddit moderators navigating threaded discussions and how it fits into their existing workflows.
In addition to evaluating the usability and practical viability of both the guidelines and resulting design features in \system{}, we identify specific features and design elements that are most and least helpful to moderators.

\subsection{Participants}
We recruited 10 moderators who are current Reddit moderators located in the United States, varied in community sizes and topics. On average, moderators had 6.5 years of experience (SD = 3.72), and the mean subreddit community size was 1.84M members (SD = 1.92M). Moderators were recruited via mod mail from large, active subreddits (top 1500 popular subreddits\footnote{Retrieved from Reddit API reddit.subreddits.popular}), and they were invited to fill out an interest form, including their moderation experience and communities they were moderating for. Participants were required to be active moderators on Reddit. We confirmed their eligibility in the interest form by verifying their moderator status via their provided username.

\subsection{Procedure}
During the screening survey, we asked moderators to indicate which subreddit(s) they currently moderate and a specific subreddit that they would like to test \system{} with. For the study sessions, the researcher collected the subreddit's data spanning two days, including posts either created or replied to within that period, to use as the database for the user study via the data collection and processing procedures outlined in Section \ref{sec:data_collection}. 

In the study, the researcher first introduced \system{}'s features. Next, moderators engaged in two primary tasks using \system{} to simulate their content moderation workflow (\S\ref{sec:task}), followed by a semi-structured interview that asked them to comment on the features and design of \system{}, as well as how \system{} can be integrated into their current workflow. Moderators were encouraged to think aloud while completing the tasks, verbalizing their decisions, questions, and reactions when they were interacting with \system{}.

All user study sessions were conducted via Zoom in English. Each session lasted approximately one hour, and each participant was compensated with a \$20 Amazon gift card. Each session was audio-recorded, transcribed, and anonymized. We also log their interaction data with \system{}.

\paragraph{Ethical Considerations} To minimize disruption to online communities, we simulated the moderation environment using scraped data collected a day prior to each user study session. This approach ensured participants worked with recent, familiar content while avoiding direct intervention in live communities. Participant recruitment was conducted through modmail, and we carefully excluded communities whose rules explicitly prohibited unsolicited messages or research recruitment. We obtained oral informed consent from all participants and implemented measures to protect anonymity by avoiding collecting real names or personally identifiable information. All potentially identifying information was removed. All research procedures were reviewed and approved by our Institutional Review Board.

\subsection{Tasks}
\label{sec:task}
To mimic a typical moderation process and to encourage the exploration of \system{}'s features, we designed the following tasks for moderators. During the study, moderators were encouraged to think aloud and describe how each feature influenced their decision-making process. If moderators did not explore certain features, researchers would reintroduce the features to ensure comprehensive evaluation.

\subsubsection{Locating Threads.} In the first task, moderators were asked to locate discussion threads of interest within their subreddit. Using \system{}’s overview features, they explored the feed to identify posts that might warrant their attention. 

\subsubsection{Moderating Comments.} In the second task, moderators clicked into one of the identified posts to review its recent comment threads. For each comment, they could decide whether to remove it if it clearly violated subreddit rules, or report it if it was ambiguous. 

\subsubsection{Interview.} Following the two tasks, each participant took part in a semi-structured interview. The interview explored their overall impressions of \system{}, how it compared to their existing moderation workflow, and how specific features supported or hindered their decision-making. Participants also provided feedback on desired improvements and reflected on \system{}’s potential to reduce cognitive load during moderation.

\subsection{Analysis}
We analyzed the interview data using thematic analysis. Given our focus on system evaluation, we deductively mapped participant feedback against our three initial Design Goals (Section \ref{sec:design-goals}) to validate \system{}'s objectives. Simultaneously, we remained open to inductive themes from the data outside of these pre-defined categories. During the coding process, two researchers first collaboratively coded three transcripts, discussing their interpretations and identifying preliminary themes, such as challenges moderators face, their needs, and how system features could address those needs. These discussions produced an initial, flexible set of codes and themes. One researcher then coded the remaining transcripts, continuing to consult with the second researcher when uncertainties arose.
\section{Results}

From our study, we found that \system{} successfully transformed design guidelines into practices and can be integrated to support moderation workflows in several key ways. All participants agreed that \system{} has the potential to aid in moderation tasks or in their own moderation processes. Reactions were predominantly positive, though some concerns emerged regarding the toxicity detection mismatches with community-specific norms. Despite these considerations, five of ten moderators had outlined ways that \system{} would complement their existing toolkit effectively (Section~\ref{sec:integration}). 

We organize our findings to reflect both \system{}'s intended design objectives and the broader sociotechnical context of moderation. We begin with moderators' overall reactions to \system{}'s interface and features (Section~\ref{sec:results_general}), which provides important context for understanding how they engaged with specific design elements. We then examine how \system{} addressed each of its three design goals (Sections~\ref{sec:result_goal1}--\ref{sec:result_goal3}), evaluating whether \system{} achieved its intended objectives. Finally, we present themes that emerged inductively regarding \system{}'s practical viability in real-world contexts: how moderators envision integrating \system{} into their existing workflows (Section~\ref{sec:integration}) and remaining challenges that need to be addressed for practical deployment (Section~\ref{sec:open_challenges}).

\subsection{Moderator Impressions of Interfaces and Features in \system{}}
\label{sec:results_general}

Overall, moderators expressed that the interface is quick and easy to use. In the feed-level interface, eight out of ten moderators expressed that they found the features involving toxicity values to be useful. They adjusted the toxicity threshold (Figure~\ref{fig:interface}-a2) more often than the score threshold, with an average 270.8 adjustments to the toxicity threshold per user study session compared to 35.6 for the score threshold. The histograms (Figure~\ref{fig:interface}-B) were found to be generally helpful for making sense of the toxicity threshold as well as for understanding the high-level community health. However, one moderator did report that the histogram was difficult to understand and that they were not familiar with this type of visualization. Seven out of ten moderators found the time setting and time zooming (zoom in/out timespan) not necessary, as they mostly moderate relatively frequently on the subreddits and do not need to adjust time range.

Participants also found the sorting effective, and many would sort by the highest toxicity comments (Figure~\ref{fig:interface}-C). In terms of each post and the accompanying visualizations, moderators found it easy to understand. For most moderators (7 out of 10), the temporal graphs (Figure~\ref{fig:interface}-d2) were useful for highlighting anomalies in threads and understanding trends; however, half of them (5 out of 10) also note that they used the bar (Figure~\ref{fig:interface}-d1) to make comparisons between posts, therefore guiding their attention to the most toxic posts. 

At the feed-level, moderators generally began with the system’s default toxicity threshold (0.2) and sorted by the highest-toxicity comments (6 out of 10), as three of them mentioned sorting is a common functionality they used in Reddit's moderation tool (ModQueue). To gain an early sense of community health and to orient themselves to the range of toxicity values, they often glanced at the visualizations for the first few surfaced posts. From there, moderators drew on the visual summaries in ways that fit their own routines: the histogram helped them interpret the overall distribution of toxicity across the subreddit, while the temporal activity charts could show toxicity spikes in problematic behavior occurred as they adjusted the threshold. Because their first task was to identify a thread to examine in depth, all moderators adjusted the threshold or sort until the charts revealed a post containing at least some toxic comments.

In the thread-level interface, moderators generally found the interface easy to navigate without confusion. The highlighting and filtering features (Figure~\ref{fig:interface-thread}-e1\&e2) were the most appreciated features identified by moderators, with eight out of ten reporting them as useful. Moderators also expressed that the approve and delete buttons were also intuitive and helpful as they showed their moderation actions on the posts.
However, due to the low number of replies most top-level comments receive, the visualizations for TLCs (Figure~\ref{fig:interface-thread}-F) were not as helpful as the visualizations for threads, as 6 out of 10 moderators noted, and the ``jumping to'' feature (Figure~\ref{fig:interface-thread}-f3\&g3) was not often used.

At the thread level, moderators were typically drawn directly to the highlighted comments or filtered to view only toxic comments. They then began examining these comments more closely, adjusting the toxicity threshold as needed once they could see how the toxicity values mapped onto the actual content.

\subsection{Design Goal 1: Transforming Textual Content into Intuitive Visual Patterns}
\label{sec:result_goal1}

We found that our multi-level visual summaries were largely effective in surfacing patterns in recent activity and supporting efficient moderator awareness.

Seven out of ten moderators reported that thread-level visualizations were effective, with three moderators specifically highlighting the effectiveness of visualizing individual TLCs. Moderators frequently expressed that our visualization approach offered superior functionality compared to existing Reddit moderation tools and has the potential to fit right into their preexisting workflows. One moderator specifically stated how \system{}'s feed-level overview aligns with their typical practices:
\begin{quote}
    \textit{``This is interesting, like, [visualizations] for threads is much more useful than what we have...in ModQueue or toolbox, you can't see that[discussions/replies] at all. So I've gone into the thread to go make sure I get everything that's in there.''} (P1)
\end{quote}

The visualization features that most effectively supported awareness and detection of problematic content were the thread-level toxicity graphs and the temporal activity visualizations. These tools transformed moderation from sequential text review to pattern recognition, allowing moderators to quickly identify anomalies without manually reviewing each comment individually. As P2 noted: 
\begin{quote}
    \textit{``If I see this graph and it starts not looking like this[toxicity line stays low], then I should do something about it. I think there's value in that, and it makes it very easy for me.''} 
\end{quote}

This capability proved especially valuable for identifying contentious discussions: 
\begin{quote}
    \textit{``If I could just look at that visualization of what's happening on the subreddit, then that could help me choose which threads to look at and see if there's too much fighting going on.''} (P7)
\end{quote}

These visual patterns may substantially alter traditional moderation workflows. Rather than processing comments one-by-one, moderators adopted a more strategic approach: first identifying potential hotspots through our visualizations, then investigating those specific areas. This triage-based workflow can support more efficient attention allocation by directing moderator attention to areas of greatest concern. Additionally, the ability to visually scan for patterns enabled more proactive moderation, with moderators identifying potentially problematic threads before they accumulated multiple reports. 

Within individual TLC visualization, moderators reported limited utility due to low amount of activity within the replies of individual TLCs. As P1 noted: 
\begin{quote}
    \textit{``I think the visualizations work pretty well for what their intended purpose is like, even with just the small number of comments you scraped like they made sense... the visualizations work. We just don't give it enough to visualize.''} 
\end{quote}

This suggests that while the visualizations function as intended even with modest data, visualization effectiveness scales with content volume, being most valuable at the thread level where patterns become more meaningful.

\subsection{Design Goal 2: Directing Attention Through Metrics and Visual Cues}
\label{sec:result_goal2}

We found that \system{}'s implementation of toxicity metrics and highlighting features was successful in supporting informed decision making for moderators. These features effectively addressed the moderators' need for attention-directing mechanisms that could help them navigate large volumes of content efficiently.

\subsubsection{Enhanced Decision Support Through Toxicity Metrics}

Moderators found \system{}'s toxicity metrics were notably more effective than Reddit's native tools. As P6 remarked:
\begin{quote}
    \textit{``[Needle] seems to work a little bit more intelligently than the tools that Reddit has initially...this is the sort of tool that can help you find that[unreported contents] for any sort of subreddits.''}
\end{quote} 

Specifically, these toxicity metrics helped moderators pinpoint problematic content that might otherwise go unnoticed, with P1 noting that it was more helpful than standard Reddit tools since the ModQueue sometimes missed toxic comments that were unreported.

The value of toxicity metrics was particularly evident for high-volume subreddits where manual review becomes impractical. P5 explained the challenge of scale in very large subreddits where moderators need to perform thousands of actions per day just to keep pace with incoming content. While our study did not measure task completion time quantitatively, four moderators mentioned that in these environments, toxicity detection significantly reduced their cognitive load to identify toxicity. Three moderators expressed interest in customizable toxicity thresholds that would allow them to receive alerts when unusual patterns emerge. P9 suggested they would set high thresholds to identify threads receiving an abnormal number of toxic comments, allowing for more targeted intervention.

Another common usage of toxicity metrics was as a reference point to aid in their decision-making process. For example, P10 said that: 
\begin{quote}
    \textit{``[Toxicity is like a] second opinion on stuff if I don't know I should remove it or not.''}
\end{quote}

\subsubsection{Improving Content Processing Through Visual Highlighting}

The highlighting feature proved particularly effective at directing moderator attention, with eight out of ten moderators acknowledging and appreciating this function primarily for its efficiency benefits. P1 noted that highlighting would help them process threaded sections much faster. And as P5 noted, quick processing is a fundamental requirement for effective moderation. The color-coded system provided immediate visual cues that guide moderator attention, with P3 explaining that their eye naturally gets drawn to content highlighted in red, which guided their attention similarly to how their existing filters flag content. 

Visual highlighting also appeared to improve moderation thoroughness, with three moderators mentioning how it helped them catch problematic content they might otherwise have missed when reviewing comments across their entire subreddit. By combining toxicity metrics with visual highlights, moderators were able to implement a more strategic, targeted approach to moderation.

Although our highlight function identifies comments based on multiple metrics—including high scores, toxicity levels, and combined metrics (toxicity and score)---six moderators specifically noted that score metrics fall outside their typical concerns in moderation process. Therefore, score is not a useful metric for them. Thus, the combination of toxicity and visual highlighting emerged as particularly effective. Together, they were described by moderators as reducing cognitive effort and enabling more systematic navigation of large comment volumes.

\subsection{Design Goal 3: Adapting to Community Needs Through Customizable Settings}
\label{sec:result_goal3}

We found that our implementation of adjustable thresholds gave moderators meaningful ways to customize \system{} to their needs. Our findings support the importance of this flexibility, particularly in how moderators interpret and apply toxicity thresholds across diverse community contexts.

Moderators actively experimented with toxicity thresholds to calibrate the tool to their community's specific communication norms. They demonstrated a process of exploratory calibration to find appropriate settings. P4 described needing to adjust thresholds repeatedly to determine the right sensitivity level for their community. By testing content against different thresholds, moderators gained additional context for judgment calls. Moderators also expressed interest in understanding the underlying mechanics of toxicity detection to better calibrate thresholds. For example, P1 noted curiosity about what specific text patterns triggered higher toxicity values.

\subsection{Integration with Existing Moderation Workflows}
\label{sec:integration}

We investigated how \system{} can integrate with existing moderation practices as it is essential for evaluating its practical utility. Moderators reported using a variety of native Reddit tools including AutoMod, ModQueue, and user history cards, while operating across both mobile and desktop platforms. Three moderators also mentioned utilizing external communication tools such as Discord and Slack for moderator discussions and coordination. Therefore, due to the diversity in moderation tools, there is no ``one-size-fits-all'' solution that \system{} can fit into every tool. 

We found that the visualization graphs could integrate into existing workflows in meaningful ways by helping moderators identify emerging problematic topics and add relevant keywords from these discussions to their filters. For example, P2 explained that:
\begin{quote}
    \textit{``If we see a big spike in that graph, we're just like, okay, what is the topic that's being discussed? Let's go ahead and add, you know, the words related to this topic to our auto moderator.''}
\end{quote} 
This demonstrates how \system{}'s visualizations can directly enhance existing AutoMod configurations by helping moderators proactively identify conversation trends and trajectories that require additional monitoring.

Toxicity metrics were also reported to be useful to integrate into existing workflows. Five moderators mentioned performing moderation activities directly from the Reddit feed, using sorting functions to prioritize content review. This behavior aligns well with \system{}'s feed-level interface and enhanced sorting capabilities (with toxicity metrics), suggesting another way to integrate the tool into existing workflows.

\subsection{Open Challenges and Unaddressed Moderator Needs}
\label{sec:open_challenges}

While \system{} demonstrated effectiveness in addressing workflow efficiency challenges, our interviews revealed a persistent challenge that requires further development: the nuanced understanding of community-specific language and norms. Six moderators consistently emphasized how community-dependent language complicates automated toxicity detection across different communities. This highlights the need for more community- and context-sensitive models to detect norm violations~\cite{jurgens2019just, goyal2025momoe}.

Moderators across gaming, sports, and fandom subreddits highlighted how domain-specific language often triggers false positives in toxicity detection. P5 (from a game's subreddit) noted that ``toxic'' phrases can be praise in gaming culture, while P2 (from a sport's subreddit) pointed to emotional sports talk as a source of misclassification. P3 (from a TV show's subreddit) and P6 (from a game's subreddit) emphasized the role of show-specific jargon and fictional context in defining toxicity. While \system{}’s adjustable thresholds help, moderators like P9 stressed the need for a model that understands niche vocabularies and cultural nuance, as well as specific removal patterns in the community. 
\section{Discussion}

In this paper, we presented a multi-tiered visual analytics approach in \system{}, translated design principles from prior HCI research into practice, and demonstrated how \system{} can enhance moderation workflows beyond Reddit’s current tools. Through a user evaluation study with 10 Reddit moderators, we found that \system{}'s design and features effectively addressed three core challenges in content moderation: (1) detecting anomalies via visual patterns in text, (2) surfacing problematic content through toxicity metrics and visual cues, and (3) accommodating diverse community norms through customizable thresholds. 

\system{} provides aggregated metrics at the subreddit level for comprehensive community health assessment, visualizes toxicity and activity patterns at the thread level to reveal conversation health without requiring exhaustive reading, and incorporates temporal visualizations to help moderators detect emerging issues. This layered approach helped transform overwhelming textual content into intuitive, scannable visualization patterns that can reduce cognitive load on Reddit moderators (G1).

We also introduced attention-directing mechanisms in \system{} that combined multiple data sources to identify potentially problematic content (G2). By integrating toxicity analysis, community signals like voting patterns, temporal anomalies that may indicate coordinated behavior, and customizable highlighting, \system{} helped moderators focus on areas requiring intervention. By synthesizing these signals, \system{} supported more informed decision-making while preserving human judgment in the moderation process.

Finally, we developed a customizable moderation system that can be adapted to diverse community needs and contexts (G3). \system{} incorporates adjustable thresholds for different metrics, visualization scaling that accommodates various discussion sizes, and customizable panels that allows moderators to tailor their view based on community-specific priorities. This flexibility ensures that the tool can serve communities with different norms, sizes, and moderation approaches without requiring technical expertise to configure.

Next, we discuss the implications of our findings, design considerations to inform future tools, and next steps for \system{}’s development.
 
\subsection{Tools to Navigate Threaded Discussions and Guide Moderator Attention}

\system{}'s success highlights the effectiveness of visualization-based approaches in transforming moderation workflows from sequential text review to pattern recognition. By converting textual content into visual patterns, moderators can quickly identify anomalies and potential issues without manually examining each comment. This approach fundamentally changes how moderators navigate large volumes of content---moving from linear processing to a more strategic workflow that improves both efficiency and effectiveness, as well as helping distance moderators from immediate trauma \cite{CaseyModerationChallenge, emotionalLabor, psycModerator}.

While prior systems like ConvEx \cite{convex} or ConVis \cite{convis} support sense-making in linear or topic-based formats, they do not address the ``branching'' complexity of threaded discussions \cite{aragon_thread_2017}. In \system{}, particularly effective was the thread-level toxicity visualization that allowed moderators to identify problematic discussions at a glance. As P2 noted, these visualizations make it ``very easy'' to determine when intervention is necessary. This capability proves especially valuable for high-volume subreddits where manual review becomes impractical, addressing what P5 described as the ``scale challenge'' where moderators need to perform thousands of actions daily.

Additionally, \system{} supports a shift from individual to community-level situational awareness. Unlike Reddit's current ModQueue, where comments are presented as isolated infractions stripped of conversational context \cite{bajpai_queue_2025}, \system{}'s visualizations serve as a signal directing moderator attention to specific, escalating threads. Furthermore, customizable thresholds allow moderators to explicitly calibrate these signals to their specific community norms \cite{thread_caution, politicsAlgo, proactive_review}, empowering them to intervene early in toxic cascades \cite{thread_caution}.

To illustrate the practical implications of \system{}, Table~\ref{tab:baseline_comparison} compares the workflows in \system{} with native moderation tools including AutoMod and ModQueue in Reddit. Reddit's ModQueue is the primary interface used by moderators, which shows a linear list of comments that are identified through AutoMod rules or flagged by users pending reviews from the moderators \cite{bajpai_queue_2025}.

\begin{table*}[t]
\centering
\small
\renewcommand{\arraystretch}{1.5} 
\begin{tabular}{p{2.5cm} p{4cm} p{4cm} p{3.5cm}}
\toprule
\textbf{Design Goal (DG)} & \textbf{Native Tools} & \textbf{\system{}'s Workflow} & \textbf{Conceptual Shift} \\
\midrule
\textbf{DG1: Multi-level Visual Summarizations} & 
Linear comment lists in ModQueue strip conversational context across threaded branches. Moderators must manually click into threads to understand discussion flow and reply hierarchies \cite{bajpai_queue_2025}. & 
Feed-level temporal visualizations and thread-level toxicity graphs enable pattern recognition at a glance. Segmented breakdown bars compress entire discussions into scannable signatures showing toxic/highly-scored content proportions. & 
Moderators transitioned from sequential text review to strategic triage, identifying hotspots before reading individual comments.  \\
\midrule
\textbf{DG2: Directing Attention \& Providing Actionable Insights} & 
Moderators rely on community reports and keyword-based AutoMod filters \cite{jhaver2019automated}. Flagged content appears isolated as individual comments in ModQueue without surrounding conversational context, potentially missing unreported violations. & 
Visualizes toxicity and voting signals in situ within the threaded conversations. Uses color-coded highlights to guide attention to problematic clusters without removing them from the surrounding dialogue. Customizable sorting prioritizes high-toxicity threads. & 
Moderators shifted moderation from reactive flagging to proactive conversation health monitoring.  \\
\midrule
\textbf{DG3: Adapting to Diverse Community Needs} & 
Static AutoMod keyword rules require manual configuration and apply uniformly once set \cite{jhaver2019automated}. While keywords can be tailored to communities, they lack nuance for context-dependent toxicity (e.g., whether gaming slang is toxic depends on conversational context, not just the words themselves) \cite{chandrasekharan2018internet}. & 
Adjustable toxicity and score thresholds with real-time histogram feedback showing community-wide distributions. Moderators calibrate sensitivity through exploratory threshold testing to match their subreddit's communication culture. & 
Moderators actively experimented with thresholds to accommodate domain-specific language compared to a rigid list of keywords. \\
\bottomrule
\end{tabular}
\caption{Conceptual comparison of moderation workflows supported by native Reddit tools and \system{}, categorized by our design goals.}
\Description{Table comparing native Reddit moderation tools and the Needle’s workflow across three design goals, multi-level visual summarization, directing attention and providing actionable insights, and adapting to diverse community needs. Each row lists a design goal, describes how native tools address it, how Needle addresses it, and the conceptual shift.}
\label{tab:baseline_comparison}
\end{table*}

\subsection{Design Implications}

We identify several key implications for designing effective tools specifically suited for community-driven moderation:

\subsubsection{Customizability and Adaptability to Community Norms}

A critical theme that emerged in our study is the need to account for community-specific norms.
Unlike commercial moderation that enforces uniform global policies \cite{psycModerator}, volunteer moderation often involves adapting to community rules and cultures \cite{chandrasekharan_crossmod_2019}. This variability creates challenges for standardized toxicity detection systems, which often fail to account for contextual nuances in community discourse. Future moderation tools should incorporate community-sensitive models that can learn and adapt to specific subreddit cultures, rules, and communication patterns.

The need for customization goes beyond adjustable toxicity thresholds to fit into different moderation philosophies. Some communities prefer tools that support positive reinforcement in addition to content removal. Four moderators, for instance, described practices such as awarding flairs, identifying potential future moderators, and extracting high-quality comments to include into community wikis. This aligns with recent HCI work exploring the role of positive reinforcement in moderation~\cite{lambert2024positive} and highlights the need for future research into systems that facilitate such positive moderation actions.

\subsubsection{Multi-level Agency and Overview}

\system{}'s implementation of both feed-level and thread-level interfaces demonstrates the importance of providing multiple perspectives on community content. Moderators need the ability to transition between high-level pattern recognition and detailed comment review. This multi-level approach supports different moderation strategies---from proactive identification of emerging issues \cite{habib2019actreactinvestigatingproactive, proactive_review} to targeted intervention in specific discussions \cite{convex}.

Effective moderation tools should provide clear visualization pathways between these levels, allowing moderators to quickly drill down from community-wide patterns to specific problematic content. This capability supports informed decision-making by providing both context and detail when needed.

\subsubsection{Human-in-the-Loop Approaches}

While automation can significantly improve moderation efficiency, our study reinforces the critical role of human judgment in content moderation. Our findings suggest that moderators do not view algorithmic signals as absolute truth but rather as one of several inputs to be triangulated with their own domain expertise.

Moderators valued \system{} not as a replacement for their decision-making but as a tool that enhances their capabilities through such as ``second opinion'' support (P10) and strategic attention direction. Specifically, when content sits in a gray area, these ``second opinions'' (e.g., toxicity metrics) help moderators perform a sanity-check on their own interpretations and gain confidence in borderline decisions. Additionally, by design, Needle is intended to assist moderators by drawing attention to potentially problematic areas rather than taking automated actions on their behalf. The toxicity metrics support strategic attention direction by filtering large volumes of content and highlighting areas that warrant closer human review. Three moderators also explicitly mentioned that all their moderation decisions are purely human-based to capture community norms and adherence to past moderation standards.

This finding suggests that moderation tools should be designed with a human-in-the-loop philosophy, where technology augments rather than replaces moderator judgment \cite{chandrasekharan_crossmod_2019, convex, nipitinthebud, halfaker2020ores}. The ideal approach combines algorithmic efficiency with human discernment---using technology to filter and highlight potential issues while preserving moderator agency in final decision-making. This human oversight is particularly important in the generative AI era, where automated tools and LLMs still struggle with nuance and false positives \cite{llm-mod, convex}. Human moderators remain necessary to correct algorithmic errors, making visualization tools like \system{} essential for scaling this human review process.

\subsubsection{Enabling Proactive Moderation}
A significant limitation of traditional moderation tools is their predominantly reactive nature, requiring moderators to respond to issues after they emerge. \system{}'s visualization capabilities demonstrate potential for shifting toward more proactive moderation approaches. By visualizing temporal patterns and discussion trends, moderators can monitor conversational trajectories and identify potentially problematic conversations earlier in their development~\cite{conversations_gone_awry, lambert2022conversational, thread_caution}, before they escalate to require intervention.

\subsection{Next steps for \system{}}

Based on our findings, we identify several promising directions for the future development of \system{}:

\subsubsection{Developing Community-Specific Toxicity Models}
Moderators consistently noted the limitations of generalized toxicity detection. Future work should develop subreddit-specific models trained on community norms, prior moderation actions, and unique community vocabulary. Additionally, incorporating contextual signals, such as discussion topics and user reputation, could help the system better interpret niche language and improve detection accuracy.

Designers and developers should collect details on the specifics about community rules and practices, as they vary in forms~\cite{fiesler2018reddit, reddy2023evolution}. For example, one moderator explicitly mentioned that they used word count to remove comments, as their subreddit required sophisticated answers citing credible sources; another moderator expressed that image-based models would be useful to detect toxicity, as a lot of toxicity came from images in their community. 

\subsubsection{User-Level Information Integration}
Five moderators emphasized the value of incorporating user history and behavior patterns into moderation decisions. Future versions of \system{} could include user-level signals such as prior comment quality or moderation history directly in the interface, helping moderators determine whether a comment is part of a broader behavioral pattern. These signals could be further visualized or calculated to similar metrics or summaries currently used for comment- and thread-level insights.

\subsubsection{Positive Moderation Features}
While \system{} currently focuses on identifying problematic content, moderators expressed interest in tools that would also help them identify and promote positive reinforcement~\cite{lambert2025does, choi2025creator}. Future development could expand \system{}'s capabilities to include features for recognizing desirable behavior~\cite{goyal2026uncovering}, highlighting quality comments~\cite{park2016supporting}, and implementing positive reinforcement through awards, flairs, highlights, or adding to the subreddit's wiki or weekly digest to incentivize participation and motivate contributions~\cite{gilbert2018motivations, lambert2024positive, lambert2025mind}.

\subsubsection{Expanding \system{} to Assist Reddit Users}
Beyond moderation, \system{} could also support everyday users in navigating and making sense of large-scale discussions. For example, visual summaries like activity bursts and toxicity spikes can help users obtain overviews of discussion threads. Similarly, surfacing highly active threads can guide users to the most relevant or interesting parts of a discussion. These features may improve engagement and promote more informed participation among social media users. 
Broadly, further research is needed to examine the influence of both textual and visual summaries on how users consume, interpret, and make sense of large-scale online discussions.

\subsection{Limitations}

The study presents several limitations.

First, the evaluation was conducted in a simulated environment with pre-scraped data rather than during live moderation. This choice minimized disruption to real communities and allowed us to control for variability in Reddit’s API, but it also meant that we were unable to capture the pressures and dynamics of real-time moderation.

Second, \system{} was evaluated as a standalone tool rather than integrated with moderators’ existing workflows. We chose this approach to isolate and examine its design features, and we asked participants to reflect on how it might fit alongside their current tools. However, this setup leaves open questions about exactly how \system{} would be adopted and used in practice when combined with tools such as AutoMod and ModQueue. We envision a practical path to adoption as a browser-based extension. This approach would allow visualizations to function as an additional layer over Reddit's current interface. 

Third, our evaluation did not include an ablation study to isolate the individual contributions of specific system components (e.g., temporal visualizations or toxicity metrics). We chose this approach because these components are tightly coupled and intended to be used together to support moderation workflows. As a result, we do not attribute observed workflow changes to individual components in isolation. As we envision \system{} working alongside existing moderation ecosystems, future longitudinal studies could examine how these features perform in combination or isolation with other moderation tools.

Fourth, regarding scalability and data retrieval, \system{} currently relies on a pre-fetched JSON cache to ensure low latency for the user study. In deployment, however, exclusively using real-time Reddit API calls would face rate limits and network delays. A more scalable approach would be to register target subreddits, pre-fetch historical conversations, and run continuous background scraping, enabling a hybrid of caching and incremental loading that balances responsiveness with data completeness.

Lastly, we focused exclusively on Reddit to examine threaded discussions. While threaded discussions also appear on other online platforms (e.g., X or other forums with similar affordances and moderation ecosystems~\cite{bajpai2022harmonizing}), we selected Reddit as a representative testbed given its large scale and deeply layered thread structures. Further research is needed to examine how visualizations can support users in navigating threaded discussions on other platforms.

\section{Conclusion}

Our system, \system{}, demonstrates how visualization-based approaches can transform content moderation workflows by converting overwhelming textual discussions into interpretable visual patterns. Our user study with 10 Reddit moderators revealed that the system's multi-level summarization capabilities, toxicity metrics, and customizable interfaces can reduce cognitive load while enabling proactive identification of problematic content. Although challenges still remain in automated toxicity detection across diverse community contexts, \system{}'s human-in-the-loop design successfully augments moderator capabilities without replacing human judgment. Our system can be readily extended to incorporate additional conversational metrics and community-specific signals. We hope that our findings will inform future approaches to creating visualization-driven tools that enhance human moderation capabilities across diverse online communities.

\begin{acks}
The authors would like to thank the participants for their time and contributions and the reviewers for their thoughtful feedback.
We also thank members of the Social Computing Lab (SCUBA) at the University of Illinois for their feedback and input on this work.
\end{acks}

\bibliographystyle{ACM-Reference-Format}
\bibliography{references}

@inproceedings{moderationVis,
  author={Vaidya, Sahaj and Cai, Jie and Basu, Soumyadeep and Naderi, Azadeh and Wohn, Donghee Yvette and Dasgupta, Aritra},
  booktitle={2021 IEEE Visualization Conference (VIS)}, 
  title={Conceptualizing Visual Analytic Interventions for Content Moderation}, 
  year={2021},
  volume={},
  number={},
  pages={191-195},
publisher = {IEEE},
  doi={10.1109/VIS49827.2021.9623288}}

@inproceedings{bajpai_queue_2025,
author = {Bajpai, Tanvi and Chandrasekharan, Eshwar},
title = {``Think about it like you're a firefighter'': Understanding How Reddit Moderators Use the Modqueue},
year = {2026},
booktitle = {Proceedings of the 2026 CHI Conference on Human Factors in Computing Systems},
series = {CHI '26}
}

@article{Lambert2024Positive,
author = {Lambert, Charlotte and Choi, Frederick and Chandrasekharan, Eshwar},
title = {"Positive reinforcement helps breed positive behavior": Moderator Perspectives on Encouraging Desirable Behavior},
year = {2024},
issue_date = {November 2024},
publisher = {Association for Computing Machinery},
address = {New York, NY, USA},
volume = {8},
number = {CSCW2},
doi = {10.1145/3686929},
journal = {Proc. ACM Hum.-Comput. Interact.},
month = nov,
articleno = {390},
numpages = {33}
}

@article{convex,
author = {Choi, Frederick and Bajpai, Tanvi and Pratipati, Sowmya and Chandrasekharan, Eshwar},
title = {ConvEx: A Visual Conversation Exploration System for Discord Moderators},
year = {2023},
issue_date = {October 2023},
publisher = {Association for Computing Machinery},
address = {New York, NY, USA},
volume = {7},
number = {CSCW2},
url = {https://doi.org/10.1145/3610053},
doi = {10.1145/3610053},
journal = {Proc. ACM Hum.-Comput. Interact.},
month = oct,
articleno = {262},
numpages = {30}
}

@inproceedings{onlineHarrLitReview,
author = {Huang, Evey Jiaxin and Sarma, Abhraneel and Hwang, Sohyeon and Chandrasekharan, Eshwar and Chancellor, Stevie},
title = {Opportunities, tensions, and challenges in computational approaches to addressing online harassment},
year = {2024},
isbn = {9798400705830},
publisher = {Association for Computing Machinery},
address = {New York, NY, USA},
url = {https://doi.org/10.1145/3643834.3661623},
doi = {10.1145/3643834.3661623},
booktitle = {Proceedings of the 2024 ACM Designing Interactive Systems Conference},
pages = {1483–1498},
numpages = {16},
location = {Copenhagen, Denmark},
series = {DIS '24}
}

@article{twiterToxic, title={Users’ Behavioral and Emotional Response to Toxicity in Twitter Conversations}, volume={18}, url={https://ojs.aaai.org/index.php/ICWSM/article/view/31295}, DOI={10.1609/icwsm.v18i1.31295}, abstractNote={Prior works have shown connections between online toxicity attacks, such as harassment, cyberbullying, and hate speech, and the subsequent increase in offline violence, as well as negative psychological effects on victims. These correlations are primarily identified through user studies conducted via virtual environments, simulations, and questionnaires. However, no work has investigated how, in practice and authentically, people react to online toxicity both emotionally, showing anger, anxiety, and sadness, and behaviorally in terms of engaging with and responding to toxicity instigators, considering conversations as a whole and the relation between emotions and behaviors. This data-driven study investigates the effect of toxicity on Twitter users’ behaviors and emotions considering confounding factors, such as account identifiability, activity, and conversation’s structure and topic. We collected about 80K Twitter conversations and identified those with and without toxic replies. Performing statistical tests along with propensity score matching, we investigated the causal association of receiving toxicity and users’ responses. We found that authors of conversations with toxic replies are more likely to engage in conversations, reply in a toxic way, and unfollow toxicity instigators. In terms of users’ emotional responses, we found that sadness and anger after the first toxic reply are more likely to increase as the amount of toxicity increases. These findings not only emphasize the negative emotional and behavioral effects of online toxicity on social media users but also, as demonstrated in this paper, can be utilized to build prediction models for users’ reactions, which could then aid the implementation of proactive detection and intervention measures helping users in such situations.}, number={1}, journal={Proceedings of the International AAAI Conference on Web and Social Media}, author={Aleksandric, Ana and Saha Roy, Sayak and Pankaj, Hanani and Wilson, Gabriela Mustata and Nilizadeh, Shirin}, year={2024}, month={May}, pages={29-42} }

@article{darkFacebook,
title = {The dark side of Facebook®: The Dark Tetrad, negative social potency, and trolling behaviours},
journal = {Personality and Individual Differences},
volume = {102},
pages = {79-84},
year = {2016},
issn = {0191-8869},
doi = {https://doi.org/10.1016/j.paid.2016.06.043},
url = {https://www.sciencedirect.com/science/article/pii/S0191886916307930},
author = {Naomi Craker and Evita March}
}

@article{journalistHarm,
author = {Na Yeon Lee and Ahran Park},
title ={How online harassment affects Korean journalists? The effects of online harassment on the journalists’ psychological problems and their intention to leave the profession},

journal = {Journalism},
volume = {25},
number = {4},
pages = {900-920},
year = {2024},
doi = {10.1177/14648849231166511}
}

@article{defineToxicity,
title = {Defining and detecting toxicity on social media: context and knowledge are key},
journal = {Neurocomputing},
volume = {490},
pages = {312-318},
year = {2022},
issn = {0925-2312},
doi = {https://doi.org/10.1016/j.neucom.2021.11.095},
url = {https://www.sciencedirect.com/science/article/pii/S0925231221018087},
author = {Amit Sheth and Valerie L. Shalin and Ugur Kursuncu}
}

@article{onlineMentalHealthModeration, title={The Effect of Moderation on Online Mental Health Conversations}, volume={15}, url={https://ojs.aaai.org/index.php/ICWSM/article/view/18100}, DOI={10.1609/icwsm.v15i1.18100}, abstractNote={Many people struggling with mental health issues are unable to access adequate care due to high costs and a shortage of mental health professionals, leading to a global mental health crisis. Online mental health communities can help mitigate this crisis by offering a scalable, easily accessible alternative to in-person sessions with therapists or support groups. However, people seeking emotional or psychological support online may be especially vulnerable to the kinds of antisocial behavior that sometimes occur in online discussions. Moderation can improve online discourse quality, but we lack an understanding of its effects on online mental health conversations.In this work, we leveraged a natural experiment, occurring across 200,000 messages from 7,000 online mental health conversations, to evaluate the effects of moderation on online mental health discussions. We found that participation in group mental health discussions led to improvements in psychological perspective, and that these improvements were larger in moderated conversations. The presence of a moderator increased user engagement, encouraged users to discuss negative emotions more candidly, and dramatically reduced bad behavior among chat participants. Moderation also encouraged stronger linguistic coordination, which is indicative of trust building. In addition, moderators who remained active in conversations were especially successful in keeping conversations on topic. Our findings suggest that moderation can serve as a valuable tool to improve the efficacy and safety of online mental health conversations. Based on these findings, we discuss implications and trade-offs involved in designing effective online spaces for mental health support.}, number={1}, journal={Proceedings of the International AAAI Conference on Web and Social Media}, author={Wadden, David and August, Tal and Li, Qisheng and Althoff, Tim}, year={2021}, month={May}, pages={751-763} }

@article{politicsAlgo,
author = {Robert Gorwa and Reuben Binns and Christian Katzenbach},
title ={Algorithmic content moderation: Technical and political challenges in the automation of platform governance},

journal = {Big Data \& Society},
volume = {7},
number = {1},
pages = {2053951719897945},
year = {2020},
doi = {10.1177/2053951719897945},
}

@article{convis,
author = {Hoque, E. and Carenini, G.},
title = {ConVis: A Visual Text Analytic System for Exploring Blog Conversations},
journal = {Computer Graphics Forum},
volume = {33},
number = {3},
pages = {221-230},
doi = {https://doi.org/10.1111/cgf.12378},
year = {2014}
}

@inproceedings{emotionalLabor,
author = {Wohn, Donghee Yvette},
title = {Volunteer Moderators in Twitch Micro Communities: How They Get Involved, the Roles They Play, and the Emotional Labor They Experience},
year = {2019},
isbn = {9781450359702},
publisher = {Association for Computing Machinery},
address = {New York, NY, USA},
url = {https://doi.org/10.1145/3290605.3300390},
doi = {10.1145/3290605.3300390},
booktitle = {Proceedings of the 2019 CHI Conference on Human Factors in Computing Systems},
pages = {1–13},
numpages = {13},
location = {Glasgow, Scotland Uk},
series = {CHI '19}
}

@article{CaseyModerationChallenge,
author = {Jiang, Jialun Aaron and Kiene, Charles and Middler, Skyler and Brubaker, Jed R. and Fiesler, Casey},
title = {Moderation Challenges in Voice-based Online Communities on Discord},
year = {2019},
issue_date = {November 2019},
publisher = {Association for Computing Machinery},
address = {New York, NY, USA},
volume = {3},
number = {CSCW},
url = {https://doi.org/10.1145/3359157},
doi = {10.1145/3359157},
journal = {Proc. ACM Hum.-Comput. Interact.},
month = nov,
articleno = {55},
numpages = {23}
}

@article{AmyFormFrom,
author = {Zhang, Amy X. and Bernstein, Michael S. and Karger, David R. and Ackerman, Mark S.},
title = {Form-From: A Design Space of Social Media Systems},
year = {2024},
issue_date = {April 2024},
publisher = {Association for Computing Machinery},
address = {New York, NY, USA},
volume = {8},
number = {CSCW1},
url = {https://doi.org/10.1145/3641006},
doi = {10.1145/3641006},
journal = {Proc. ACM Hum.-Comput. Interact.},
month = apr,
articleno = {167},
numpages = {47}
}

@INPROCEEDINGS{1509067,
  author={Craft, B. and Cairns, P.},
  booktitle={Ninth International Conference on Information Visualisation (IV'05)}, 
  title={Beyond guidelines: what can we learn from the visual information seeking mantra?}, 
  year={2005},
  volume={},
  number={},
  pages={110-118},
  doi={10.1109/IV.2005.28}
}

@inproceedings{synthesizedSocialSignals,
author = {Im, Jane and Tandon, Sonali and Chandrasekharan, Eshwar and Denby, Taylor and Gilbert, Eric},
title = {Synthesized Social Signals: Computationally-Derived Social Signals from Account Histories},
year = {2020},
isbn = {9781450367080},
publisher = {Association for Computing Machinery},
address = {New York, NY, USA},
url = {https://doi.org/10.1145/3313831.3376383},
doi = {10.1145/3313831.3376383},
booktitle = {Proceedings of the 2020 CHI Conference on Human Factors in Computing Systems},
pages = {1–12},
numpages = {12},
location = {Honolulu, HI, USA},
series = {CHI '20}
}

@inproceedings{AmyWikum,
author = {Zhang, Amy X. and Verou, Lea and Karger, David},
title = {Wikum: Bridging Discussion Forums and Wikis Using Recursive Summarization},
year = {2017},
isbn = {9781450343350},
publisher = {Association for Computing Machinery},
address = {New York, NY, USA},
url = {https://doi.org/10.1145/2998181.2998235},
doi = {10.1145/2998181.2998235},
booktitle = {Proceedings of the 2017 ACM Conference on Computer Supported Cooperative Work and Social Computing},
pages = {2082–2096},
numpages = {15},
location = {Portland, Oregon, USA},
series = {CSCW '17}
}

@article{AmyTilda,
author = {Zhang, Amy X. and Cranshaw, Justin},
title = {Making Sense of Group Chat through Collaborative Tagging and Summarization},
year = {2018},
issue_date = {November 2018},
publisher = {Association for Computing Machinery},
address = {New York, NY, USA},
volume = {2},
number = {CSCW},
url = {https://doi.org/10.1145/3274465},
doi = {10.1145/3274465},
journal = {Proc. ACM Hum.-Comput. Interact.},
month = nov,
articleno = {196},
numpages = {27}
}

@inproceedings{psycModerator,
author = {Steiger, Miriah and Bharucha, Timir J and Venkatagiri, Sukrit and Riedl, Martin J. and Lease, Matthew},
title = {The Psychological Well-Being of Content Moderators: The Emotional Labor of Commercial Moderation and Avenues for Improving Support},
year = {2021},
isbn = {9781450380966},
publisher = {Association for Computing Machinery},
address = {New York, NY, USA},
url = {https://doi.org/10.1145/3411764.3445092},
doi = {10.1145/3411764.3445092},
booktitle = {Proceedings of the 2021 CHI Conference on Human Factors in Computing Systems},
articleno = {341},
numpages = {14},
location = {Yokohama, Japan},
series = {CHI '21}
}

@article{chandrasekharan_crossmod_2019,
    title = {Crossmod: {A} {Cross}-{Community} {Learning}-based {System} to {Assist} {Reddit} {Moderators}},
    volume = {3},
    issn = {2573-0142},
    shorttitle = {Crossmod},
    url = {https://dl.acm.org/doi/10.1145/3359276},
    doi = {10.1145/3359276},
    language = {en},
    number = {CSCW},
    urldate = {2024-08-28},
    journal = {Proceedings of the ACM on Human-Computer Interaction},
    author = {Chandrasekharan, Eshwar and Gandhi, Chaitrali and Mustelier, Matthew Wortley and Gilbert, Eric},
    month = nov,
    year = {2019},
    pages = {1--30},
}

@article{nipitinthebud,
author = {Hsieh, Jane and Kim, Joselyn and Dabbish, Laura and Zhu, Haiyi},
title = {"Nip it in the Bud": Moderation Strategies in Open Source Software Projects and the Role of Bots},
year = {2023},
issue_date = {October 2023},
publisher = {Association for Computing Machinery},
address = {New York, NY, USA},
volume = {7},
number = {CSCW2},
url = {https://doi.org/10.1145/3610092},
doi = {10.1145/3610092},
journal = {Proc. ACM Hum.-Comput. Interact.},
month = oct,
articleno = {301},
numpages = {29}
}

@article{Park2016TopicDrift,
  author    = {Park, Albert and Hartzler, Andrea L. and Huh, Jina and Hsieh, Gary and McDonald, David W. and Pratt, Wanda},
  title     = {{``How Did We Get Here?''}: Topic Drift in Online Health Discussions},
  journal   = {Journal of Medical Internet Research},
  volume    = {18},
  number    = {11},
  pages     = {e284},
  year      = {2016},
  publisher = {JMIR Publications},
  doi       = {10.2196/jmir.6297},
  url       = {https://doi.org/10.2196/jmir.6297},
  pmid      = {27806924}
}

@INPROCEEDINGS{kang_thread,
  author={Kang, Jeon-Hyung and Kim, Jihie},
  booktitle={2011 IEEE Third International Conference on Privacy, Security, Risk and Trust and 2011 IEEE Third International Conference on Social Computing}, 
  title={Analyzing Answers in Threaded Discussions Using a Role-Based Information Network}, 
  year={2011},
  volume={},
  number={},
  pages={111-117},
  doi={10.1109/PASSAT/SocialCom.2011.107}}

@INPROCEEDINGS{Kerr_email_thread,
  author={Kerr, B.},
  booktitle={IEEE Symposium on Information Visualization 2003 (IEEE Cat. No.03TH8714)}, 
  title={Thread Arcs: an email thread visualization}, 
  year={2003},
  volume={},
  number={},
  pages={211-218},
  doi={10.1109/INFVIS.2003.1249028}}

@inproceedings{nakikj2017disvis,
  title     = {DisVis: Visualizing Discussion Threads in Online Health Communities},
  author    = {Nakikj, Drashko and Mamykina, Lena},
  booktitle = {AMIA Annual Symposium Proceedings},
  year      = {2017},
  volume    = {2016},
  pages     = {944--953},
  publisher = {American Medical Informatics Association}
}

@inproceedings{llm-mod,
author = {Kolla, Mahi and Salunkhe, Siddharth and Chandrasekharan, Eshwar and Saha, Koustuv},
title = {LLM-Mod: Can Large Language Models Assist Content Moderation?},
year = {2024},
isbn = {9798400703317},
publisher = {Association for Computing Machinery},
address = {New York, NY, USA},
url = {https://doi.org/10.1145/3613905.3650828},
doi = {10.1145/3613905.3650828},
booktitle = {Extended Abstracts of the CHI Conference on Human Factors in Computing Systems},
articleno = {217},
numpages = {8},
location = {Honolulu, HI, USA},
series = {CHI EA '24}
}

@inproceedings{goyal2025momoe,
  title={MoMoE: Mixture of Moderation Experts Framework for AI-Assisted Online Governance},
  author={Goyal, Agam and Zhan, Xianyang and Chen, Yilun and Saha, Koustuv and Chandrasekharan, Eshwar}, 
 booktitle={Proceedings of the 2025 Conference on Empirical Methods in Natural Language Processing},
  year={2025}
}

@inproceedings{gurjar2025argcmv,
  title={ArgCMV: An Argument Summarization Benchmark for the LLM-era},
  author={Gurjar, Omkar and Goyal, Agam and Chandrasekharan, Eshwar},
  booktitle={Proceedings of the 2025 Conference on Empirical Methods in Natural Language Processing},
  year={2025}
}

@article{bajpai2022harmonizing,
author = {Bajpai, Tanvi and Asher, Drshika and Goswami, Anwesa and Chandrasekharan, Eshwar},
title = {Harmonizing the Cacophony with MIC: An Affordance-aware Framework for Platform Moderation},
year = {2022},
issue_date = {November 2022},
publisher = {Association for Computing Machinery},
address = {New York, NY, USA},
volume = {6},
number = {CSCW2},
url = {https://doi-org.proxy2.library.illinois.edu/10.1145/3555110},
doi = {10.1145/3555110},
journal = {Proc. ACM Hum.-Comput. Interact.},
month = nov,
articleno = {385},
numpages = {22}
}

@article{goyal2026uncovering,
  title={Uncovering the Internet's Hidden Values: An Empirical Study of Desirable Behavior Using Highly-Upvoted Content on Reddit},
  author={Goyal, Agam and Lambert, Charlotte and Jain, Yoshee and Chandrasekharan, Eshwar},
journal = {Proceedings of the International AAAI Conference on Web and Social Media},
  year={2026},
}

@article{chan2025examining,
  title={Examining Algorithmic Curation on Social Media: An Empirical Audit of Reddit's r/popular Feed},
  author={Chan, Jackie and Choi, Fred and Saha, Koustuv and Chandrasekharan, Eshwar},
  journal = {Proceedings of the International AAAI Conference on Web and Social Media},
  year={2026}
}

@inproceedings{lambert2022conversational,
  title={Conversational resilience: Quantifying and predicting conversational outcomes following adverse events},
  author={Lambert, Charlotte and Rajagopal, Ananya and Chandrasekharan, Eshwar},
  booktitle={Proceedings of the International AAAI Conference on Web and Social Media},
  volume={16},
  pages={548--559},
  year={2022}
}

@inproceedings{gilbert2018motivations,
author = {Gilbert, Sarah},
title = {Motivations for Participating in Online Initiatives: Exploring Contributory Behaviour Across Initiative Types},
year = {2017},
isbn = {9781450346887},
publisher = {Association for Computing Machinery},
address = {New York, NY, USA},
doi = {10.1145/3022198.3024941},
booktitle = {Companion of the 2017 ACM Conference on Computer Supported Cooperative Work and Social Computing},
pages = {65–68},
numpages = {4},
location = {Portland, Oregon, USA},
series = {CSCW '17 Companion}
}

@inproceedings{park2016supporting,
author = {Park, Deokgun and Sachar, Simranjit and Diakopoulos, Nicholas and Elmqvist, Niklas},
title = {Supporting Comment Moderators in Identifying High Quality Online News Comments},
year = {2016},
isbn = {9781450333627},
publisher = {Association for Computing Machinery},
address = {New York, NY, USA},
url = {https://doi-org.proxy2.library.illinois.edu/10.1145/2858036.2858389},
doi = {10.1145/2858036.2858389},
booktitle = {Proceedings of the 2016 CHI Conference on Human Factors in Computing Systems},
pages = {1114–1125},
numpages = {12},
location = {San Jose, California, USA},
series = {CHI '16}
}

@inproceedings{zhan2024slm,
    title = "{SLM}-Mod: Small Language Models Surpass {LLM}s at Content Moderation",
    author = "Zhan, Xianyang  and
      Goyal, Agam  and
      Chen, Yilun  and
      Chandrasekharan, Eshwar  and
      Saha, Koustuv",
    booktitle = "Proceedings of the 2025 Conference of the Nations of the Americas Chapter of the Association for Computational Linguistics: Human Language Technologies (Volume 1: Long Papers)",
    month = apr,
    year = "2025",
    address = "Albuquerque, New Mexico",
    publisher = "Association for Computational Linguistics",
    url = "https://aclanthology.org/2025.naacl-long.441/",
    doi = "10.18653/v1/2025.naacl-long.441",
    pages = "8774--8790",
    ISBN = "979-8-89176-189-6"
}

@misc{lambert2025mind,
      title={Mind Your Ps and Qs: Supporting Positive Reinforcement in Moderation Through a Positive Queue}, 
      author={Charlotte Lambert and Agam Goyal and Eunice Mok and Eshwar Chandrasekharan},
      year={2025},
      eprint={2509.18437},
      archivePrefix={arXiv},
      primaryClass={cs.HC},
      url={https://arxiv.org/abs/2509.18437}, 
}

@article{thread_caution,
author = {Chang, Jonathan P. and Schluger, Charlotte and Danescu-Niculescu-Mizil, Cristian},
title = {Thread With Caution: Proactively Helping Users Assess and Deescalate Tension in Their Online Discussions},
year = {2022},
issue_date = {November 2022},
publisher = {Association for Computing Machinery},
address = {New York, NY, USA},
volume = {6},
number = {CSCW2},
url = {https://doi.org/10.1145/3555603},
doi = {10.1145/3555603},
journal = {Proc. ACM Hum.-Comput. Interact.},
month = nov,
articleno = {545},
numpages = {37}
}

@inproceedings{conversations_gone_awry,
    title = "Conversations Gone Awry: Detecting Early Signs of Conversational Failure",
    author = "Zhang, Justine  and
      Chang, Jonathan  and
      Danescu-Niculescu-Mizil, Cristian  and
      Dixon, Lucas  and
      Hua, Yiqing  and
      Taraborelli, Dario  and
      Thain, Nithum",
    editor = "Gurevych, Iryna  and
      Miyao, Yusuke",
    booktitle = "Proceedings of the 56th Annual Meeting of the Association for Computational Linguistics (Volume 1: Long Papers)",
    month = jul,
    year = "2018",
    address = "Melbourne, Australia",
    publisher = "Association for Computational Linguistics",
    url = "https://aclanthology.org/P18-1125/",
    doi = "10.18653/v1/P18-1125",
    pages = "1350--1361"
}

@article{chan2024understanding,
author = {Chan, Jackie and Lambert, Charlotte and Choi, Frederick and Chancellor, Stevie and Chandrasekharan, Eshwar},
year = {2024},
month = {05},
pages = {227-240},
title = {Understanding Community Resilience: Quantifying the Effects of Sudden Popularity via Algorithmic Curation},
volume = {18},
journal = {Proceedings of the International AAAI Conference on Web and Social Media},
doi = {10.1609/icwsm.v18i1.31310}
}

@inproceedings{chan2022community,
author = {Chan, Jackie and Atreyasa, Aditi and Chancellor, Stevie and Chandrasekharan, Eshwar},
title = {Community Resilience: Quantifying the Disruptive Effects of Sudden Spikes in Activity within Online Communities},
year = {2022},
isbn = {9781450391566},
publisher = {Association for Computing Machinery},
address = {New York, NY, USA},
doi = {10.1145/3491101.3519813},
booktitle = {Extended Abstracts of the 2022 CHI Conference on Human Factors in Computing Systems},
articleno = {339},
numpages = {8},
location = {New Orleans, LA, USA},
series = {CHI EA '22}
}

@misc{habib2019actreactinvestigatingproactive,
      title={To Act or React: Investigating Proactive Strategies For Online Community Moderation}, 
      author={Hussam Habib and Maaz Bin Musa and Fareed Zaffar and Rishab Nithyanand},
      year={2019},
      eprint={1906.11932},
      archivePrefix={arXiv},
      primaryClass={cs.SI},
      url={https://arxiv.org/abs/1906.11932}, 
}

@article{proactive_review,
author = {Schluger, Charlotte and Chang, Jonathan P. and Danescu-Niculescu-Mizil, Cristian and Levy, Karen},
title = {Proactive Moderation of Online Discussions: Existing Practices and the Potential for Algorithmic Support},
year = {2022},
issue_date = {November 2022},
publisher = {Association for Computing Machinery},
address = {New York, NY, USA},
volume = {6},
number = {CSCW2},
url = {https://doi.org/10.1145/3555095},
doi = {10.1145/3555095},
journal = {Proc. ACM Hum.-Comput. Interact.},
month = nov,
articleno = {370},
numpages = {27}
}

@inproceedings{lambert2025does,
author = {Lambert, Charlotte and Saha, Koustuv and Chandrasekharan, Eshwar},
title = {Does Positive Reinforcement Work?: A Quasi-Experimental Study of the Effects of Positive Feedback on Reddit},
year = {2025},
isbn = {9798400713941},
publisher = {Association for Computing Machinery},
address = {New York, NY, USA},
doi = {10.1145/3706598.3713830},
booktitle = {Proceedings of the 2025 CHI Conference on Human Factors in Computing Systems},
articleno = {1213},
numpages = {16},
location = {
},
series = {CHI '25}
}

@inproceedings{choi2025creator,
author = {Choi, Frederick and Lambert, Charlotte and Koshy, Vinay and Pratipati, Sowmya and Do, Tue and Chandrasekharan, Eshwar},
title = {Creator Hearts: Investigating the Impact Positive Signals from YouTube Creators in Shaping Comment Section Behavior},
year = {2025},
isbn = {9798400713941},
publisher = {Association for Computing Machinery},
address = {New York, NY, USA},
doi = {10.1145/3706598.3713521},
booktitle = {Proceedings of the 2025 CHI Conference on Human Factors in Computing Systems},
articleno = {1212},
numpages = {18},
location = {
},
series = {CHI '25}
}

@article{community_rules,
author = {J. Nathan Matias },
title = {Preventing harassment and increasing group participation through social norms in 2,190 online science discussions},
journal = {Proceedings of the National Academy of Sciences},
volume = {116},
number = {20},
pages = {9785-9789},
year = {2019},
doi = {10.1073/pnas.1813486116},
URL = {https://www.pnas.org/doi/abs/10.1073/pnas.1813486116}
}

@article{koshy2024veniremachinelearningguidedpanel,
author = {Koshy, Vinay and Choi, Frederick and Chiang, Yi-Shyuan and Sundaram, Hari and Chandrasekharan, Eshwar and Karahalios, Karrie},
title = {Venire: A Machine Learning-Guided Panel Review System for Community Content Moderation},
year = {2025},
issue_date = {November 2025},
publisher = {Association for Computing Machinery},
address = {New York, NY, USA},
volume = {9},
number = {7},
url = {https://doi-org.proxy2.library.illinois.edu/10.1145/3757699},
doi = {10.1145/3757699},
journal = {Proc. ACM Hum.-Comput. Interact.},
month = oct,
articleno = {CSCW518},
numpages = {35}
}

@inproceedings{10.1145/2998181.2998277,
author = {Seering, Joseph and Kraut, Robert and Dabbish, Laura},
title = {Shaping Pro and Anti-Social Behavior on Twitch Through Moderation and Example-Setting},
year = {2017},
isbn = {9781450343350},
publisher = {Association for Computing Machinery},
address = {New York, NY, USA},
url = {https://doi.org/10.1145/2998181.2998277},
doi = {10.1145/2998181.2998277},
booktitle = {Proceedings of the 2017 ACM Conference on Computer Supported Cooperative Work and Social Computing},
pages = {111–125},
numpages = {15},
location = {Portland, Oregon, USA},
series = {CSCW '17}
}

@techreport{inserra2024guide,
  title        = {A Guide to Content Moderation for Policymakers},
  author       = {David Inserra},
  institution  = {Cato Institute},
  type         = {Policy Analysis},
  number       = {974},
  year         = {2024},
  month        = {May},
  address      = {Washington, DC},
  url          = {https://www.cato.org/policy-analysis/guide-content-moderation-policymakers}
}

@Article{	  grimmelmann2015virtues,
  title		= {The virtues of moderation},
  author	= {Grimmelmann, James},
  journal	= {Yale JL \& Tech.},
  volume	= {17},
  pages		= {42},
  year		= {2015},
  publisher	= {HeinOnline}
}

@Book{		  roberts2019behind,
  title		= {Behind the Screen: Content Moderation in the Shadows of
		  Social Media},
  author	= {Roberts, Sarah T},
  year		= {2019},
  publisher	= {Yale University Press}
}

@Misc{		  glaser2018,
  author	= {Glaser, April},
  booktitle	= {Slate},
  title		= {{Want a Terrible Job? Facebook and Google May Be
		  Hiring.}},
  url		= {https://slate.com/technology/2018/01/facebook-and-google-are-building-an-army-of-content-moderators-for-2018.html},
  urldate	= {2018-03-27},
  year		= {2018}
}

@Misc{		  madrigal2018,
  author	= {Madrigal, Alexis C.},
  booktitle	= {The Atlantic},
  title		= {{Inside Facebook's Fast-Growing Content-Moderation
		  Effort}},
  url		= {https://www.theatlantic.com/technology/archive/2018/02/what-facebook-told-insiders-about-how-it-moderates-posts/552632/},
  urldate	= {2018-03-27},
  year		= {2018}
}

@inproceedings{matias2018civilservant,
author = {Matias, J. Nathan and Mou, Merry},
title = {CivilServant: Community-Led Experiments in Platform Governance},
year = {2018},
isbn = {9781450356206},
publisher = {Association for Computing Machinery},
address = {New York, NY, USA},
doi = {10.1145/3173574.3173583},
booktitle = {Proceedings of the 2018 CHI Conference on Human Factors in Computing Systems},
pages = {1–13},
numpages = {13},
location = {Montreal QC, Canada},
series = {CHI '18}
}

@InCollection{	  taylor2018watch,
  title		= "Regulating the networked broadcasting frontier",
  booktitle	= "Watch me play: Twitch and the rise of game live streaming",
  author	= "Taylor, TL",
  chapter	= 5,
  year		= 2018,
  publisher	= "Princeton University Press"
}

@article{seering2019moderator,
author = {Joseph Seering and Tony Wang and Jina Yoon and Geoff Kaufman},
title ={Moderator engagement and community development in the age of algorithms},

journal = {New Media \& Society},
volume = {21},
number = {7},
pages = {1417-1443},
year = {2019},
doi = {10.1177/1461444818821316}
}

@article{li2022all, title={All That’s Happening behind the Scenes: Putting the Spotlight on Volunteer Moderator Labor in Reddit}, volume={16}, url={https://ojs.aaai.org/index.php/ICWSM/article/view/19317}, DOI={10.1609/icwsm.v16i1.19317}, abstractNote={Online volunteers are an uncompensated yet valuable labor force for many social platforms. For example, volunteer content moderators perform a vast amount of labor to maintain online communities. However, as social platforms like Reddit favor revenue generation and user engagement, moderators are under-supported to manage the expansion of online communities. To preserve these online communities, developers and researchers of social platforms must account for and support as much of this labor as possible. In this paper, we quantitatively characterize the publicly visible and invisible actions taken by moderators on Reddit, using a unique dataset of private moderator logs for 126 subreddits and over 900 moderators. Our analysis of this dataset reveals the heterogeneity of moderation work across both communities and moderators. Moreover, we find that analyzing only visible work – the dominant way that moderation work has been studied thus far – drastically underestimates the amount of human moderation labor on a subreddit. We discuss the implications of our results on content moderation research and social platforms.}, number={1}, journal={Proceedings of the International AAAI Conference on Web and Social Media}, author={Li, Hanlin and Hecht, Brent and Chancellor, Stevie}, year={2022}, month={May}, pages={584-595} }

@article{fiesler2018reddit, title={Reddit Rules! Characterizing an Ecosystem of Governance}, volume={12}, url={https://ojs.aaai.org/index.php/ICWSM/article/view/15033}, DOI={10.1609/icwsm.v12i1.15033}, abstractNote={ &lt;p&gt; The social sharing and news aggregation site Reddit provides a unique example of an ecosystem of community-created rules. Not only do individual subreddits create and enforce their own regulations, but site-wide guidelines and norms may also influence behavior. This paper reports on a mixed-methods study of 100,000 subreddits and their rules. Our findings characterize the types of rules across Reddit, the frequency of rules at scale, and patterns of rules based on subreddit characteristics. We find that rules appear to be context-dependent for individual subreddits but also share common characteristics across the site. Taken together, our findings provide a rich description of this ecosystem of rules, motivating further inquiry into underlying mechanisms for rule formation and enforcement in online communities. &lt;/p&gt; }, number={1}, journal={Proceedings of the International AAAI Conference on Web and Social Media}, author={Fiesler, Casey and Jiang, Jialun and McCann, Joshua and Frye, Kyle and Brubaker, Jed}, year={2018}, month={Jun.} }

@inproceedings{jurgens2019just,
    title = "A Just and Comprehensive Strategy for Using {NLP} to Address Online Abuse",
    author = "Jurgens, David  and
      Hemphill, Libby  and
      Chandrasekharan, Eshwar",
    editor = "Korhonen, Anna  and
      Traum, David  and
      M{\`a}rquez, Llu{\'i}s",
    booktitle = "Proceedings of the 57th Annual Meeting of the Association for Computational Linguistics",
    month = jul,
    year = "2019",
    address = "Florence, Italy",
    publisher = "Association for Computational Linguistics",
    url = "https://aclanthology.org/P19-1357/",

    pages = "3658--3666",
}

@inproceedings{reddy2023evolution,
author = {Reddy, Harita and Chandrasekharan, Eshwar},
title = {Evolution of Rules in Reddit Communities},
year = {2023},
isbn = {9798400701290},
publisher = {Association for Computing Machinery},
address = {New York, NY, USA},
doi = {10.1145/3584931.3606973},
booktitle = {Companion Publication of the 2023 Conference on Computer Supported Cooperative Work and Social Computing},
pages = {278–282},
numpages = {5},
location = {Minneapolis, MN, USA},
series = {CSCW '23 Companion}
}

@article{halfaker2020ores,
author = {Halfaker, Aaron and Geiger, R. Stuart},
title = {ORES: Lowering Barriers with Participatory Machine Learning in Wikipedia},
year = {2020},
issue_date = {October 2020},
publisher = {Association for Computing Machinery},
address = {New York, NY, USA},
volume = {4},
number = {CSCW2},
url = {https://doi-org.proxy2.library.illinois.edu/10.1145/3415219},
doi = {10.1145/3415219},
journal = {Proc. ACM Hum.-Comput. Interact.},
month = oct,
articleno = {148},
numpages = {37}
}

@inproceedings{kiene2019volunteer,
author = {Kiene, Charles and Grandprey-Shores, Kate and Chandrasekharan, Eshwar and Jhaver, Shagun and Jiang, Jialun "Aaron" and Dym, Brianna and Seering, Joseph and Gilbert, Sarah and Lo, Kat and Wohn, Donghee Yvette and Dosono, Bryan},
title = {Volunteer Work: Mapping the Future of Moderation Research},
year = {2019},
isbn = {9781450366922},
publisher = {Association for Computing Machinery},
address = {New York, NY, USA},
doi = {10.1145/3311957.3359443},
booktitle = {Companion Publication of the 2019 Conference on Computer Supported Cooperative Work and Social Computing},
pages = {492–497},
numpages = {6},
location = {Austin, TX, USA},
series = {CSCW '19 Companion}
}

@article{cai2021after,
author = {Cai, Jie and Wohn, Donghee Yvette},
title = {After Violation But Before Sanction: Understanding Volunteer Moderators' Profiling Processes Toward Violators in Live Streaming Communities},
year = {2021},
issue_date = {October 2021},
publisher = {Association for Computing Machinery},
address = {New York, NY, USA},
volume = {5},
number = {CSCW2},
url = {https://doi-org.proxy2.library.illinois.edu/10.1145/3479554},
doi = {10.1145/3479554},
journal = {Proc. ACM Hum.-Comput. Interact.},
month = oct,
articleno = {410},
numpages = {25}
}

@article{chandrasekharan2018internet,
author = {Chandrasekharan, Eshwar and Samory, Mattia and Jhaver, Shagun and Charvat, Hunter and Bruckman, Amy and Lampe, Cliff and Eisenstein, Jacob and Gilbert, Eric},
title = {The Internet's Hidden Rules: An Empirical Study of Reddit Norm Violations at Micro, Meso, and Macro Scales},
year = {2018},
issue_date = {November 2018},
publisher = {Association for Computing Machinery},
address = {New York, NY, USA},
volume = {2},
number = {CSCW},
url = {https://doi-org.proxy2.library.illinois.edu/10.1145/3274301},
doi = {10.1145/3274301},
journal = {Proc. ACM Hum.-Comput. Interact.},
month = nov,
articleno = {32},
numpages = {25}
}

@inproceedings{saha2019prevalence,
author = {Saha, Koustuv and Chandrasekharan, Eshwar and De Choudhury, Munmun},
title = {Prevalence and Psychological Effects of Hateful Speech in Online College Communities},
year = {2019},
isbn = {9781450362023},
publisher = {Association for Computing Machinery},
address = {New York, NY, USA},
doi = {10.1145/3292522.3326032},
booktitle = {Proceedings of the 10th ACM Conference on Web Science},
pages = {255–264},
numpages = {10},
location = {Boston, Massachusetts, USA},
series = {WebSci '19}
}

@Book{		  gillespie2018custodians,
  title		= {Custodians of the Internet: Platforms, content moderation,
		  and the hidden decisions that shape social media},
  author	= {Gillespie, Tarleton},
  year		= {2018},
  publisher	= {Yale University Press}
}

@article{jhaver2019automated,
author = {Jhaver, Shagun and Birman, Iris and Gilbert, Eric and Bruckman, Amy},
 title = {Human-Machine Collaboration for Content Regulation: The Case of Reddit Automoderator},
 journal = {ACM Trans. Comput.-Hum. Interact.},
 issue_date = {July 2019},
 volume = {26},
 number = {5},
 month = jul,
 year = {2019},
 issn = {1073-0516},
 pages = {31:1--31:35},
 articleno = {31},
 numpages = {35},
 url = {http://doi.acm.org/10.1145/3338243},
 doi = {10.1145/3338243},
 acmid = {3338243},
 publisher = {ACM},
}

@article{jhaver2019explanations,
author = {Jhaver, Shagun and Bruckman, Amy and Gilbert, Eric},
title = {Does Transparency in Moderation Really Matter? User Behavior After Content Removal Explanations on Reddit},
year = {2019},
issue_date = {November 2019},
publisher = {Association for Computing Machinery},
address = {New York, NY, USA},
volume = {3},
number = {CSCW},
doi = {10.1145/3359252},
journal = {Proc. ACM Hum.-Comput. Interact.},
month = nov,
articleno = {150},
numpages = {27}
}

@inproceedings{10.1145/3584931.3607499,
author = {Liao, Jingxian and Singh, Mrinalini and Hung, Yi-Ting and Lin, Wen-Chieh and Wang, Hao-Chuan},
title = {ConceptCombo: Assisting Online Video Access with Concept Mapping and Social Commenting Visualizations},
year = {2023},
isbn = {9798400701290},
publisher = {Association for Computing Machinery},
address = {New York, NY, USA},
url = {https://doi.org/10.1145/3584931.3607499},
doi = {10.1145/3584931.3607499},
booktitle = {Companion Publication of the 2023 Conference on Computer Supported Cooperative Work and Social Computing},
pages = {362–364},
numpages = {3},
location = {Minneapolis, MN, USA},
series = {CSCW '23 Companion}
}

@inproceedings{10.1145/3678884.3681821,
author = {Brannon, William and Beeferman, Doug and Jiang, Hang and Heyward, Andrew and Roy, Deb},
title = {AudienceView: AI-Assisted Interpretation of Audience Feedback in Journalism},
year = {2024},
isbn = {9798400711145},
publisher = {Association for Computing Machinery},
address = {New York, NY, USA},
url = {https://doi.org/10.1145/3678884.3681821},
doi = {10.1145/3678884.3681821},
booktitle = {Companion Publication of the 2024 Conference on Computer-Supported Cooperative Work and Social Computing},
pages = {65–68},
numpages = {4},
location = {San Jose, Costa Rica},
series = {CSCW Companion '24}
}

@article{budak_threading_2017,
    title = {Threading is {Sticky}: {How} {Threaded} {Conversations} {Promote} {Comment} {System} {User} {Retention}},
    volume = {1},
    issn = {2573-0142},
    shorttitle = {Threading is {Sticky}},
    url = {https://dl.acm.org/doi/10.1145/3134662},
    doi = {10.1145/3134662},
    language = {en},
    number = {CSCW},
    urldate = {2025-08-24},
    journal = {Proceedings of the ACM on Human-Computer Interaction},
    author = {Budak, Ceren and Garrett, R. Kelly and Resnick, Paul and Kamin, Julia},
    month = dec,
    year = {2017},
    pages = {1--20},
}

@article{hadfi_structural_2024,
    title = {Structural complexity predicts consensus readability in online discussions},
    volume = {14},
    issn = {1869-5469},
    url = {https://link.springer.com/10.1007/s13278-024-01212-1},
    doi = {10.1007/s13278-024-01212-1},
    language = {en},
    number = {1},
    urldate = {2025-08-24},
    journal = {Social Network Analysis and Mining},
    author = {Hadfi, Rafik and Ito, Takayuki},
    month = mar,
    year = {2024},
    pages = {51},
}

@inproceedings{guo_throwaway_2025,
    address = {San Diego, CA, USA},
    title = {Throwaway {Accounts} and {Moderation} on {Reddit}},
    isbn = {979-8-9919276-5-9},
    url = {https://www.ndss-symposium.org/wp-content/uploads/usec25-31.pdf},
    doi = {10.14722/usec.2025.23031},
    language = {en},
    urldate = {2025-08-25},
    booktitle = {Proceedings 2025 {Symposium} on {Usable} {Security}},
    publisher = {Internet Society},
    author = {Guo, Cheng and Caine, Kelly},
    year = {2025},
}

@article{graham_sociomateriality_2021,
    title = {The {Sociomateriality} of {Rating} and {Ranking} {Devices} on {Social} {Media}: {A} {Case} {Study} of {Reddit}’s {Voting} {Practices}},
    volume = {7},
    issn = {2056-3051},
    shorttitle = {The {Sociomateriality} of {Rating} and {Ranking} {Devices} on {Social} {Media}},
    url = {https://doi.org/10.1177/20563051211047667},
    doi = {10.1177/20563051211047667},
    language = {EN},
    number = {3},
    urldate = {2025-08-25},
    journal = {Social Media + Society},
    author = {Graham, Timothy and Rodriguez, Aleesha},
    month = jul,
    year = {2021},
    note = {Publisher: SAGE Publications Ltd},
    pages = {20563051211047667},
}

@inproceedings{lampe_slashdot_2004,
    address = {Vienna Austria},
    title = {Slash(dot) and burn: distributed moderation in a large online conversation space},
    isbn = {978-1-58113-702-6},
    shorttitle = {Slash(dot) and burn},
    url = {https://dl.acm.org/doi/10.1145/985692.985761},
    doi = {10.1145/985692.985761},
    language = {en},
    urldate = {2025-08-25},
    booktitle = {Proceedings of the {SIGCHI} {Conference} on {Human} {Factors} in {Computing} {Systems}},
    publisher = {ACM},
    author = {Lampe, Cliff and Resnick, Paul},
    month = apr,
    year = {2004},
    pages = {543--550},
}

@inproceedings{vargo_deciding_2024,
  title     = {Deciding to Delete Posts on Reddit: What Factors Influence Content Removal},
  author    = {Vargo, Chris and Masullo, Gina and Hopp, Tobias},
  booktitle = {Proceedings of the 57th Hawaii International Conference on System Sciences},
  pages     = {2593--2602},
  year      = {2024},
  address   = {Hawaii},
  publisher = {ScholarSpace, University of Hawai'i at Manoa},
  doi       = {10.24251/HICSS.2024.314},
  url       = {https://hdl.handle.net/10125/106696}
}

@article{cima_taming_2024,
title = {Investigating the heterogeneous effects of a massive content moderation intervention via Difference-in-Differences},
journal = {Online Social Networks and Media},
volume = {48},
pages = {100320},
year = {2025},
issn = {2468-6964},
doi = {https://doi.org/10.1016/j.osnem.2025.100320},
url = {https://www.sciencedirect.com/science/article/pii/S2468696425000217},
author = {Lorenzo Cima and Benedetta Tessa and Amaury Trujillo and Stefano Cresci and Marco Avvenuti}
}

@article{blumer_tracking_2025,
    title = {Tracking patterns in toxicity and antisocial behavior over user lifetimes on large social media platforms},
    volume = {15},
    issn = {2045-2322},
    url = {https://www.ncbi.nlm.nih.gov/pmc/articles/PMC12260000/},
    doi = {10.1038/s41598-025-07086-3},
    urldate = {2025-08-25},
    journal = {Scientific Reports},
    author = {Blumer, Katy and Kleinberg, Jon},
    month = jul,
    year = {2025},
    pmid = {40659657},
    pmcid = {PMC12260000},
    pages = {25369},
}

@article{aragon_thread_2017,
    title = {To {Thread} or {Not} to {Thread}: {The} {Impact} of {Conversation} {Threading} on {Online} {Discussion}},
    volume = {11},
    issn = {2334-0770, 2162-3449},
    shorttitle = {To {Thread} or {Not} to {Thread}},
    url = {https://ojs.aaai.org/index.php/ICWSM/article/view/14880},
    doi = {10.1609/icwsm.v11i1.14880},
    language = {en},
    number = {1},
    urldate = {2025-05-20},
    journal = {Proceedings of the International AAAI Conference on Web and Social Media},
    author = {Aragón, Pablo and Gómez, Vicenç and Kaltenbrunner, Andreaks},
    month = may,
    year = {2017},
    pages = {12--21},
}

@article{kiene_technological_2019,
    title = {Technological {Frames} and {User} {Innovation}: {Exploring} {Technological} {Change} in {Community} {Moderation} {Teams}},
    volume = {3},
    issn = {2573-0142},
    shorttitle = {Technological {Frames} and {User} {Innovation}},
    url = {https://dl.acm.org/doi/10.1145/3359146},
    doi = {10.1145/3359146},
    language = {en},
    number = {CSCW},
    urldate = {2025-09-05},
    journal = {Proceedings of the ACM on Human-Computer Interaction},
    author = {Kiene, Charles and Jiang, Jialun Aaron and Hill, Benjamin Mako},
    month = nov,
    year = {2019},
    pages = {1--23},
}

@INPROCEEDINGS{shneiderman_eyes_nodate,
  author={Shneiderman, B.},
  booktitle={Proceedings 1996 IEEE Symposium on Visual Languages}, 
  title={The eyes have it: a task by data type taxonomy for information visualizations}, 
  year={1996},
  volume={},
  number={},
  pages={336-343},
  doi={10.1109/VL.1996.545307}}

@article{crawford_what_2016,
    title = {What is a flag for? {Social} media reporting tools and the vocabulary of complaint},
    volume = {18},
    issn = {1461-4448, 1461-7315},
    shorttitle = {What is a flag for?},
    url = {https://journals.sagepub.com/doi/10.1177/1461444814543163},
    doi = {10.1177/1461444814543163},
    language = {en},
    number = {3},
    urldate = {2025-09-05},
    journal = {New Media \& Society},
    author = {Crawford, Kate and Gillespie, Tarleton},
    month = mar,
    year = {2016},
    pages = {410--428},
}

@INPROCEEDINGS{horne_identifying_2017,
  author={Horne, Benjamin D. and Adali, Sibel and Sikdar, Sujoy},
  booktitle={2017 26th International Conference on Computer Communication and Networks (ICCCN)}, 
  title={Identifying the Social Signals That Drive Online Discussions: A Case Study of Reddit Communities}, 
  year={2017},
  volume={},
  number={},
  pages={1-9},
  doi={10.1109/ICCCN.2017.8038388}}

@misc{perspective_model_card,
  title        = {Perspective API Model Cards},
  author       = {{Google}},
year = {2026},
  howpublished = {\url{https://developers.perspectiveapi.com/s/about-the-api-model-cards?language=en_US}},
}

@inproceedings{lees_new_2022,
    address = {Washington DC USA},
    title = {A {New} {Generation} of {Perspective} {API}: {Efficient} {Multilingual} {Character}-level {Transformers}},
    isbn = {978-1-4503-9385-0},
    shorttitle = {A {New} {Generation} of {Perspective} {API}},
    url = {https://dl.acm.org/doi/10.1145/3534678.3539147},
    doi = {10.1145/3534678.3539147},
    language = {en},
    urldate = {2025-11-19},
    booktitle = {Proceedings of the 28th {ACM} {SIGKDD} {Conference} on {Knowledge} {Discovery} and {Data} {Mining}},
    publisher = {ACM},
    author = {Lees, Alyssa and Tran, Vinh Q. and Tay, Yi and Sorensen, Jeffrey and Gupta, Jai and Metzler, Donald and Vasserman, Lucy},
    month = aug,
    year = {2022},
    pages = {3197--3207},
}

@inproceedings{yang_towards_2023,
    address = {Singapore},
    title = {Towards {Detecting} {Contextual} {Real}-{Time} {Toxicity} for {In}-{Game} {Chat}},
    url = {https://aclanthology.org/2023.findings-emnlp.663/},
    doi = {10.18653/v1/2023.findings-emnlp.663},
    urldate = {2025-11-19},
    booktitle = {Findings of the {Association} for {Computational} {Linguistics}: {EMNLP} 2023},
    publisher = {Association for Computational Linguistics},
    author = {Yang, Zachary and Grenon-Godbout, Nicolas and Rabbany, Reihaneh},
    editor = {Bouamor, Houda and Pino, Juan and Bali, Kalika},
    month = dec,
    year = {2023},
    pages = {9894--9906},
}

@inproceedings{raman_centering_2023,
    address = {Singapore},
    title = {Centering the {Margins}: {Outlier}-{Based} {Identification} of {Harmed} {Populations} in {Toxicity} {Detection}},
    shorttitle = {Centering the {Margins}},
    url = {https://aclanthology.org/2023.emnlp-main.579/},
    doi = {10.18653/v1/2023.emnlp-main.579},
    urldate = {2025-11-19},
    booktitle = {Proceedings of the 2023 {Conference} on {Empirical} {Methods} in {Natural} {Language} {Processing}},
    publisher = {Association for Computational Linguistics},
    author = {Raman, Vyoma and Fleisig, Eve and Klein, Dan},
    editor = {Bouamor, Houda and Pino, Juan and Bali, Kalika},
    month = dec,
    year = {2023},
    pages = {9316--9329},
}

@inproceedings{zheng_hatemoderate_2024,
    address = {Mexico City, Mexico},
    title = {{HateModerate}: {Testing} {Hate} {Speech} {Detectors} against {Content} {Moderation} {Policies}},
    shorttitle = {{HateModerate}},
    url = {https://aclanthology.org/2024.findings-naacl.172/},
    doi = {10.18653/v1/2024.findings-naacl.172},
    urldate = {2025-11-19},
    booktitle = {Findings of the {Association} for {Computational} {Linguistics}: {NAACL} 2024},
    publisher = {Association for Computational Linguistics},
    author = {Zheng, Jiangrui and Liu, Xueqing and Haque, Mirazul and Qian, Xing and Yang, Guanqun and Yang, Wei},
    editor = {Duh, Kevin and Gomez, Helena and Bethard, Steven},
    month = jun,
    year = {2024},
    pages = {2691--2710},
}

@article{saveski_social_2021,
    title = {Social {Catalysts}: {Characterizing} {People} {Who} {Spark} {Conversations} {Among} {Others}},
    volume = {5},
    issn = {2573-0142},
    shorttitle = {Social {Catalysts}},
    url = {https://dl.acm.org/doi/10.1145/3476023},
    doi = {10.1145/3476023},
    language = {en},
    number = {CSCW2},
    urldate = {2025-11-19},
    journal = {Proceedings of the ACM on Human-Computer Interaction},
    author = {Saveski, Martin and Kooti, Farshad and Morelli Vitousek, Sylvia and Diuk, Carlos and Bartlett, Bryce and Adamic, Lada A.},
    month = oct,
    year = {2021},
    pages = {1--20},
}

@article{yu_signals_2025,
    title = {Signals in the {Noise}: {Decoding} {Unexpected} {Engagement} {Patterns} on {Twitter}},
    volume = {9},
    issn = {2573-0142},
    shorttitle = {Signals in the {Noise}},
    url = {https://dl.acm.org/doi/10.1145/3757657},
    doi = {10.1145/3757657},
    language = {en},
    number = {7},
    urldate = {2025-11-19},
    journal = {Proceedings of the ACM on Human-Computer Interaction},
    author = {Yu, Yulin and Chen, Houming and Romero, Daniel and Dhillon, Paramveer S.},
    month = oct,
    year = {2025},
    pages = {1--33},
}

@article{yu_characterizing_2024,
    title = {Characterizing the {Structure} of {Online} {Conversations} {Across} {Reddit}},
    volume = {8},
    issn = {2573-0142},
    url = {https://dl.acm.org/doi/10.1145/3686913},
    doi = {10.1145/3686913},
    language = {en},
    number = {CSCW2},
    urldate = {2025-11-19},
    journal = {Proceedings of the ACM on Human-Computer Interaction},
    author = {Yu, Yulin and Jiang, Julie and Dhillon, Paramveer S.},
    month = nov,
    year = {2024},
    pages = {1--23},
}

@article{
noauthor_spreading_nodate,
author = {Michela Del Vicario  and Alessandro Bessi  and Fabiana Zollo  and Fabio Petroni  and Antonio Scala  and Guido Caldarelli  and H. Eugene Stanley  and Walter Quattrociocchi },
title = {The spreading of misinformation online},
journal = {Proceedings of the National Academy of Sciences},
volume = {113},
number = {3},
pages = {554-559},
year = {2016},
doi = {10.1073/pnas.1517441113}}

@inproceedings{cheng_anyone_2017,
    address = {Portland Oregon USA},
    title = {Anyone {Can} {Become} a {Troll}: {Causes} of {Trolling} {Behavior} in {Online} {Discussions}},
    isbn = {978-1-4503-4335-0},
    shorttitle = {Anyone {Can} {Become} a {Troll}},
    url = {https://dl.acm.org/doi/10.1145/2998181.2998213},
    doi = {10.1145/2998181.2998213},
    language = {en},
    urldate = {2025-11-19},
    booktitle = {Proceedings of the 2017 {ACM} {Conference} on {Computer} {Supported} {Cooperative} {Work} and {Social} {Computing}},
    publisher = {ACM},
    author = {Cheng, Justin and Bernstein, Michael and Danescu-Niculescu-Mizil, Cristian and Leskovec, Jure},
    month = feb,
    year = {2017},
    pages = {1217--1230},
}

@inproceedings{saveski_structure_2021,
    address = {Ljubljana Slovenia},
    title = {The {Structure} of {Toxic} {Conversations} on {Twitter}},
    isbn = {978-1-4503-8312-7},
    url = {https://dl.acm.org/doi/10.1145/3442381.3449861},
    doi = {10.1145/3442381.3449861},
    language = {en},
    urldate = {2025-11-19},
    booktitle = {Proceedings of the {Web} {Conference} 2021},
    publisher = {ACM},
    author = {Saveski, Martin and Roy, Brandon and Roy, Deb},
    month = apr,
    year = {2021},
    pages = {1086--1097},
}

@inproceedings{diakopoulos_towards_2011,
    address = {Hangzhou China},
    title = {Towards quality discourse in online news comments},
    isbn = {978-1-4503-0556-3},
    url = {https://dl.acm.org/doi/10.1145/1958824.1958844},
    doi = {10.1145/1958824.1958844},
    language = {en},
    urldate = {2025-11-19},
    booktitle = {Proceedings of the {ACM} 2011 conference on {Computer} supported cooperative work},
    publisher = {ACM},
    author = {Diakopoulos, Nicholas and Naaman, Mor},
    month = mar,
    year = {2011},
    pages = {133--142},
}

\appendix
\newpage
\section{Toxicity and Score Values for Selected Comments}
\label{sec:appendix_examples}

Table~\ref{tab:comment_examples} presents a selection of comments from our user studies alongside their toxicity and score values. To protect Reddit user privacy and prevent the amplification of harmful content, all comment text has been paraphrased to preserve the original intent and tone while ensuring they are no longer searchable. To maintain the validity of these examples, we re-evaluated the paraphrased text using the Perspective API (consistent with our main study) to ensure they retained similar toxicity levels ($\pm 0.2$). We reported toxicity values for the paraphrased comments, while retaining the original score values from the source comments.

\begin{table}[ht]
\centering
\small
\renewcommand{\arraystretch}{1.4}
\begin{tabular}{p{5cm} c c}
\toprule
\textbf{Paraphrased Comment Text} & \textbf{Toxicity} & \textbf{Score} \\
\midrule
The saw is a capacitance-based sensor, not a metal detector. You can think of it like a touch screen. Can usually saw through a small nail without the safety mechanism activating. & 0.01 & 31 \\
\midrule
What’s your issue? Are you busy being a Jack ass to learn how to respond to a simple question? & 0.66 & 1 \\
\midrule
Missed times to say Oregon turns into charred duck everytime they get close to a natty & 0.39 & -10 \\
\bottomrule
\end{tabular}
\caption{Paraphrased comments displaying their Toxicity (0.0--1.0) and Score.}
\Description{Table showing three paraphrased Reddit comments alongside their toxicity scores and vote scores (referred to as scores in the paper).}
\label{tab:comment_examples}
\end{table}

\end{document}